\documentclass[preprint2,longabstract]{aastex}
\usepackage{multicol}

\newcommand{\MJ}{$M_{\rm{J}}$}
\newcommand{\Msun}{$M_{\odot}$}
\newcommand{\Rsun}{$R_{\odot}$}
\newcommand{\Rearth}{$R_{\oplus}$}
\newcommand{\To}{$t_{\rm{0}}$}
\newcommand{\uo}{$u_{\rm{0}}$}
\newcommand{\uopos}{$u_{\rm{0}} > 0$}
\newcommand{\uoneg}{$u_{\rm{0}} < 0$}
\newcommand{\TE}{$t_{\rm{E}}$}
\newcommand{\thetaE}{$\theta_{\rm{E}}$}
\newcommand{\thetaS}{$\theta_{S}$}
\newcommand{\alf}{$\alpha_{0}$}

\newcommand{\s}{$s_{0}$}
\newcommand{\q}{$q$}

\newcommand{\piee}{$\pi_{{\rm E},E}$}
\newcommand{\pien}{$\pi_{{\rm E},N}$}
\newcommand{\topar}{$t_{0,\rm{par}}$}
\newcommand{\chisq}{$\chi^{2}$}
\newcommand{\dchisq}{$\Delta\chi^{2}$}
\newcommand{\redchisq}{$\chi_{\rm{red}}^{2}$}
\newcommand{\dsdt}{$ds/dt$}
\newcommand{\dadt}{$d\alpha/dt$}
\newcommand{\fluxs}{$f_{s}$}
\newcommand{\fluxb}{$f_{b}$}

\newcommand{\DL}{$D_{L}$}
\newcommand{\DS}{$D_{S}$}
\newcommand{\RS}{$R_{S}$}
\newcommand{\MLp}{$M_{L,1}$}
\newcommand{\MLs}{$M_{L,2}$}
\newcommand{\MLtot}{$M_{L,\rm{tot}}$}
\newcommand{\hjdp}{HJD$^\prime$}
\newcommand{\VSinst}{$V_{S,\rm{inst}}$}
\newcommand{\ISinst}{$I_{S,\rm{inst}}$}
\newcommand{\VISinst}{$(V-I)_{S,\rm{inst}}$}
\newcommand{\VSo}{$V_{S,0}$}
\newcommand{\ISo}{$I_{S,0}$}
\newcommand{\VISo}{$(V-I)_{S,0}$}
\newcommand{\VKSo}{$(V-K)_{S,0}$}
\newcommand{\KSo}{$K_{S,0}$}

\newcommand{\IRCinst}{$I_{\rm{RC},\rm{inst}}$}
\newcommand{\VIRCinst}{$(V-I)_{\rm{RC},\rm{inst}}$}

\newcommand{\IRCo}{$I_{\rm{RC},\rm{0}}$}
\newcommand{\VIRCo}{$(V-I)_{\rm{RC},\rm{0}}$}
\newcommand{\IRCapp}{$I_{RC,\rm{app}}$}
\newcommand{\myemail}{rstreet@lcogt.net}

\newcommand{\thetaEvale}{0.557$\pm$0.09}
\newcommand{\thetaSvale}{9.143$\pm$0.792}
\newcommand{\RSvale}{14.7$\pm$1.3}
\newcommand{\MLpvale}{0.16$\pm$0.03}
\newcommand{\MLsvale}{11.0$\pm$2.0}
\newcommand{\MLtotvale}{0.17$\pm$0.03}
\newcommand{\DLvale}{2.8$\pm$0.4}
\newcommand{\DRC}{7.48}
\newcommand{\Ro}{8.20}
\newcommand{\DeltaIval}{0.20}
\newcommand{\aperpvale}{1.21$\pm$0.16}
\newcommand{\Eratio}{0.079}
\newcommand{\PM}{4.60$\pm$0.4}
\newcommand{\ISinstvale}{15.554$\pm$0.007}
\newcommand{\VSinstvale}{15.335$\pm$0.007}
\newcommand{\VISinstvale}{-0.22$\pm$0.01}
\newcommand{\IRCinstvale}{15.821$\pm$0.05}
\newcommand{\VIRCinstvale}{-0.350$\pm$0.05}
\newcommand{\IRCoval}{14.443}

\newcommand{\MVRCoval}{0.97}
\newcommand{\IRCappval}{14.24}

\newcommand{\VIRCoval}{1.09}
\newcommand{\ISoval}{13.976}
\newcommand{\VSoval}{15.197}
\newcommand{\VISovale}{1.221$\pm$0.051}
\newcommand{\VKSoval}{2.852}
\newcommand{\KSoval}{12.345}
\newcommand{\twoMJvale}{13.686$\pm$0.053}
\newcommand{\twoMHvale}{12.926$\pm$0.057}
\newcommand{\twoMKvale}{12.642$\pm$0.054}

\newcommand{\TEvale}{44.3$\pm$0.1}
\newcommand{\TEval}{44.3}

\newcommand{\qvale}{0.0654$\pm$0.0006}

\newcommand{\Slopevale}{-0.0153$\pm$0.0004}

\newcommand{\SSlopevale}{-0.0160$\pm$0.0004}

\newcommand{\chisqS}{6331.950}
\newcommand{\chisqSS}{5334.378}
\newcommand{\chisqSPpos}{6064.773}
\newcommand{\chisqSPOpos}{5544.071}
\newcommand{\chisqSPOSpos}{4782.367}
\newcommand{\chisqSPneg}{6099.487}
\newcommand{\chisqSPOneg}{5690.003}
\newcommand{\chisqSPOSneg}{4802.606}

\newcommand{\dchisqS}{1549.583}
\newcommand{\dchisqSS}{552.011}
\newcommand{\dchisqSP}{1282.406}
\newcommand{\dchisqSPO}{761.704}
\newcommand{\dchisqSPOS}{0.0}

\shorttitle{MOA-2010-BLG-073Lb}
\shortauthors{Street et al.}

\begin{document}

\title{MOA-2010-BLG-073L: An M-Dwarf with a Substellar Companion at the
Planet/Brown Dwarf Boundary}

\author{
R.A.~Street$^{1}$,
J.-Y.~Choi$^{2}$, 
Y.~Tsapras$^{1,3}$, 
C.~Han$^{2,\dagger}$,
K.~Furusawa$^{4}$,
M.~Hundertmark$^{5}$, 
A.~Gould$^{6}$,
T.~Sumi$^{7}$,
I.A.~Bond$^{8}$, 
D.~Wouters$^{9}$,
R.~Zellem$^{10}$,
A.~Udalski$^{11}$\\
and\\
(The RoboNet Collaboration)\\
C.~Snodgrass$^{12}$, 
K.~Horne$^{5}$,
M.~Dominik$^{5,\ddagger}$,
P.~Browne$^{5}$, 
N.~Kains$^{13}$,
D.M.~Bramich$^{13}$, 
D.~Bajek$^{5}$,
I.A.~Steele$^{14}$,\\
S.~Ipatov$^{15}$, 
and\\
(The MOA Collaboration)\\
F.~Abe$^{4}$, 
D.P.~Bennett$^{16}$, 
C.S.~Botzler$^{17}$, 
P.~Chote$^{18}$,
M.~Freeman$^{17}$,
A.~Fukui$^{4}$, 
P.~Harris$^{18}$,
Y.~Itow$^{4}$, 
C.H.~Ling$^{8}$, 
K.~Masuda$^{4}$, 
Y.~Matsubara$^{4}$, 
N.~Miyake$^{4}$,
Y.~Muraki$^{19}$,
T.~Nagayama$^{20}$, 
S.~Nishimaya$^{21}$,
K.~Ohnishi$^{22}$, 
N.~Rattenbury$^{17}$, 
To.~Saito$^{23}$,
D.J.~Sullivan$^{18}$,
D.~Suzuki$^{7}$, 
W.L.~Sweatman$^{8}$, 
P.J.~Tristram$^{24}$,
K.~Wada$^{7}$, 
P.C.M.~Yock$^{17}$\\
and\\
(The OGLE Collaboration)\\
M.\,K. Szyma{\'n}ski$^{11}$, M. Kubiak$^{11}$, G.
Pietrzy{\'n}ski$^{11,25}$, I. Soszy{\'n}ski$^{11}$, R. Poleski$^{11}$, K.
Ulaczyk$^{11}$, {\L}. Wyrzykowski$^{11}$
and\\
(The $\mu$FUN Collaboration)\\
J.~Yee$^{6}$, 
S.~Dong$^{26}$,
I.-G.~Shin$^{2}$, 
C.-U.~Lee$^{27}$, 
J.~Skowron$^{6}$, 
L.~Andrade De Almeida$^{28}$, 
D.L.~DePoy$^{29}$,
B.S.~Gaudi$^{6}$, 
L.-W.~Hung$^{6}$, 
F.~Jablonski$^{28}$,
S.~Kaspi$^{30}$, 
N.~Klein$^{30}$, 
K.-H.~Hwang$^{2}$,
J.-R.~Koo$^{27}$, 
D.~Maoz$^{30}$, 
J.A.~Mu{\~n}oz$^{31}$,
R.W.~Pogge$^{6}$, 
D.~Polishhook$^{30}$,
A.~Shporer$^{29,1}$, 
J.~McCormick$^{32}$,
G.~Christie$^{33}$,
T.~Natusch$^{33}$,
B.~Allen$^{34}$,
J.Drummond$^{35}$,
D.~Moorhouse$^{36}$,
G.~Thornley$^{36}$,
M.~Knowler$^{36}$,
M.~Bos$^{37}$,
G.~Bolt$^{38}$,
\\
and\\
(The PLANET Collaboration)\\
J.-P.~Beaulieu$^{9}$, 
M.D.~Albrow$^{39}$,
V.~Batista$^{6}$, 
S.~Brillant$^{40}$,
J.A.R.~Caldwell$^{41}$, 
A.~Cassan$^{42}$,
A.~Cole$^{43}$, 
E.~Corrales$^{42}$,
Ch.~Coutures$^{42}$, 
S.~Dieters$^{43}$,
D.~Dominis Prester$^{44}$, 
J.~Donatowicz$^{45}$, 
P.~Fouqu{\'e}$^{46,47}$,
E.~Bachelet$^{46,47}$,
J.~Greenhill$^{43}$, 
S.R.~Kane$^{48}$, 
D.~Kubas$^{47}$, 
J.-B.~Marquette$^{42}$,
R.~Martin$^{49}$, 
J.~Menzies$^{50}$,
K.R.~Pollard$^{17}$, 
K.C.~Sahu$^{51}$,
J.~Wambsganss$^{52}$, 
A.~Williams$^{49}$,
M.~Zub$^{52}$\\
and\\
(MiNDSTEp)\\
K.A.~Alsubai$^{15}$, 
V.~Bozza$^{53}$,
M.J.~Burgdorf$^{54}$, 
S.~Calchi Novati$^{55}$, 
P.~Dodds$^{5}$,
S.~Dreizler$^{56}$, 
F.~Finet$^{57}$, 
T.~Gerner$^{52}$,
S.~Hardis$^{58}$, 
K.~Harps{\o}e$^{59}$, 
F.~Hessman$^{56}$,
T.C.~Hinse$^{27}$,
U.G.~J{\o}rgensen$^{58}$,
E.~Kerins$^{60}$, 
C.~Liebig$^{5}$,
L.~Mancini$^{61,53,55}$, 
M.~Mathiasen$^{58}$,
M.T.~Penny$^{6,59}$, 
S.~Proft$^{52}$, 
S.~Rahvar$^{62}$,
D.~Ricci$^{57}$, 
G.~Scarpetta$^{63}$, 
S.~Sch{\"a}fer$^{56}$,
F.~Sch{\"o}nebeck$^{52}$, 
J.~Southworth$^{64}$,
J.~Surdej$^{57}$
}
\affil{$^{1}$LCOGT, 6740 Cortona Drive, Suite 102, Goleta, CA 93117, USA.}
\affil{$^{2}$Department of Physics, Institute for Astrophysics, 
Chungbuk National University, Cheongju, 361-763, Korea}
\affil{$^{3}$School of Mathematical Sciences, Queen Mary, 
University of London, Mile End Road, London E1 4NS, UK}
\affil{$^{4}$Solar-Terrestrial Environment Laboratory, Nagoya University, 
Nagoya, 464-8601, Japan}
\affil{$^{5}$SUPA/St~Andrews, Dept. of Physics and Astronomy, North Haugh, 
St.~Andrews, Fife, KY16~9SS, UK}
\affil{$^{6}$Dept. of Astronomy, Ohio State University, McPherson Laboratory, 
140 W.~18th Avenue, Columbus, Ohio 43210-1173, USA}
\affil{$^{7}$Dept. of Earth and Space Science, Graduate School of Science, 
Osaka University, 1-1 Machikaneyama-cho, Toyonaka, Osaka 560-0043, Japan}
\affil{$^{8}$Institute of Information and Mathematical Sciences, Massey 
University, Private Bag 102-904, North Shore Mail Centre, Auckland, New Zealand}
\affil{$^{9}$UPMC-CNRS, UMR 7095, Institut d'Astrophysique de Paris, 98bis 
boulevard Arago, F-75014, Paris, France}
\affil{$^{10}$Lunar and Planetary Laboratory,  Department of Planetary Sciences,
University of Arizona, 1629 E. University Blvd., Tucson AZ 85721-0092, USA}
\affil{$^{11}$Warsaw University Observatory, Al.~Ujazdowskie~4, 00-478~Warszawa,
Poland}
\affil{$^{12}$Max Planck Institute for Solar System Research,
Max-Planck-Str. 2, 37191 Katlenburg-Lindau, Germany}
\affil{$^{13}$European Southern Observatory, Karl-Schwarzschild-Stra{\ss}e 2, 
85748 Garching bei M\"{u}nchen, Germany}
\affil{$^{14}$Astrophysics Research Institute, Liverpool John Moores University,
Twelve Quays House, Egerton Wharf, Birkenhead, Wirral., CH41 1LD, UK}
\affil{$^{15}$Qatar Foundation, P.O.~Box 5825, Doha, Qatar}
\affil{$^{16}$Dept. of Physics, University of Notre Dame, Notre Dame, IN~46556, 
USA}
\affil{$^{17}$Dept. of Physics, University of Auckland, Private Bag~92019, 
Auckland, New Zealand}
\affil{$^{18}$School of Chemical and Physical Sciences, Victoria University, 
Wellington, New Zealand}
\affil{$^{19}$Dept. of Physics, Konan University, Nishiokamoto 8-9-1, Kobe 
658-8501, Japan}
\affil{$^{20}$Dept. of Physics and Astrophysics, Faculty of Science, Nagoya
University, Nagoya 464-8602, Japan}
\affil{$^{21}$Extrasolar Planet Detection Project Office, National Astronomical 
Observatory of Japan (NAOJ), Osawa 2-12-1, Mitaka, Tokyo 181-8588, Japan}
\affil{$^{22}$Nagano National College of Technology, Nagano, 381-8550, Japan}
\affil{$^{23}$Tokyo Metropolitan College of Industrial Technology, Tokyo, 
116-8523, Japan}
\affil{$^{24}$Mt.~John Observatory, P.O.~Box~56, Lake Tekapo 8770, New Zealand}
\affil{$^{25}$Universidad de Concepci{\'o}n, Departamento de Astronomia,
Casilla 160--C, Concepci{\'o}n, Chile}
\affil{$^{26}$Institute for Advanced Study, Einstein Drive, Princeton, New Jersey 08540, USA}
\affil{$^{27}$Korea Astronomy \& Space Science Institute, Daejeon, 305-348, 
Korea}
\affil{$^{28}$Divisao de Astrofisica, Instituto Nacional de Pesquisas 
Espeaciais, Avenida dos Astronauntas, 1758 Sao Jos{\'e} dos Campos, 12227-010 
SP, Brazil}
\affil{$^{29}$Dept. of Physics and Astronomy, Texas A\&M University College 
Station, TX 77843-4242, USA}
\affil{$^{30}$School of Physics and Astronomy and Wise Observatory, Tel-Aviv 
University, Tel-Aviv 69978, Israel}
\affil{$^{31}$Departamento de Astronomi{\'a} y Astrof{\'i}sica, Universidad de 
Valencia, E-46100 Burjassot, Valencia, Spain}
\affil{$^{32}$Farm Cove Observatory, Farm Cove, Pakuranga, Auckland 2010, New Zealand}
\affil{$^{33}$Auckland Observatory, Auckland, New Zealand}
\affil{$^{34}$Vintage Lane Observatory, Blenheim, New Zealand}
\affil{$^{35}$Possum Observatory, Patutahi, New Zealand}
\affil{$^{36}$Kumeu Observatory, Kumeu, West Auckland, New Zealand}
\affil{$^{37}$Molehill Astronomical Observatory, North Shore City, New Zealand}
\affil{$^{38}$Craigie, Perth, Western Australia, Australia}
\affil{$^{39}$University of Canterbury, Dept. of Physics and Astronomy, 
Private Bag~4800, Christchurch 8020, New Zealand}
\affil{$^{40}$European Southern Observatory, Casilla 19001, Vitacura 19, 
Santiago, Chile}
\affil{$^{41}$McDonald Observatory, 16120 St Hwy Spur 78 \#2, Fort Davis, 
Tx 79734, USA}
\affil{$^{42}$Institut d'Astrophysique de Paris, UMR7095 CNRS-Universit{\'e} 
Pierre \& Marie Curie, 98 bis boulevard Arago, 75014 Paris, France}
\affil{$^{43}$School of Math and Physics, University of Tasmania, Private Bag 
37, GPO Hobart, Tasmania 7001, Australia}
\affil{$^{44}$Physics Department, Faculty of Arts and Sciences, University of 
Rijeka, Omladinska 14, 51000 Rijeka, Croatia}
\affil{$^{45}$Technical University of Vienna, Department of Computing, Wiedner 
Hauptstrasse 10, Vienna, Austria}
\affil{$^{46}$Universit{\'e} de Toulouse, UPS-OMP, IRAP, Toulouse, France}
\affil{$^{47}$CNRS, IRAP, 14 Avenue Edouard Belin, F-31400 Toulouse, France}
\affil{$^{48}$NASA Exoplanet Science Institute, Caltech, MS 100-22, 770 
S.~Wilson Ave., Pasadena, CA 91125, USA}
\affil{$^{49}$Perth Observatory, Walnut Road, Bickley, Perth 6076, Australia}
\affil{$^{50}$South African Astronomical Observatory, P.O.~Box 9, Observatory 
7935, South Africa}
\affil{$^{51}$Space Telescope Science Institute, 3700 San Martin Drive,
Baltimore, MD 21218, USA}
\affil{$^{52}$Astronomisches Rechen-Institut, Zentrum f{\"u}r Astronomie der 
Universit{\"a}t Heidelberg (ZAH), M{\"o}nchhofstr. 12-14, 69120 Heidelberg, 
Germany}
\affil{$^{53}$Universit{\'a} degli Studi di Salerno, Dipartimento di Fisica 
``E.R. Caianiello'', Via S.~Allende, 84081, Baronissi (SA), Italy}
\affil{$^{54}$SOFIA Science Center, NASA Ames Research Center, Mail Stop 
N211-3, Moffett Field, CA 94035, USA}
\affil{$^{55}$Istituto Internazionale per gli Alti Studi Scientifici (IIASS), 
Vietri Sul Mare (SA) Italy}
\affil{$^{56}$Institut f{\"u}r Astrophysik, Georg-August-Universit{\"a}t, 
Friedrich-Hund-Platz1, 37077 G{\"o}ttingen, Germany}
\affil{$^{57}$Institut d'Astrophysique et de G{\'e}ophysique, All{\'e}e du 6 
Ao{\^u}t 17, Sart Tilman, B{\^a}t. B5c, 4000 Li{\'e}ge, Belgium}
\affil{$^{58}$Niels Bohr Institute, University of Copenhagen, Juliane 
Maries vej 30, 2100 Copenhagen, Denmark}
\affil{$^{59}$Centre for Star and Planet Formation, Geological Museum, 
{\O}ster Voldgade 5, 1350 Copenhagen, Denmark}
\affil{$^{60}$Jodrell Bank Centre for Astrophysics, University of Manchester, 
Oxford Road, Manchester, M13 9PL, UK}
\affil{$^{61}$Max Planck Institute for Astronomy, K{\"o}nigstuhl 17, 619117 
Heidelberg, Germany}
\affil{$^{62}$Dept. of Physics, Sharif University of Technology, P.O.~Box 
11155-9161, Tehran, Iran}
\affil{$^{63}$INFN, Gruppo Collegato di Salerno, Sezione di Napoli, Italy}
\affil{$^{64}$Astrophysics Group, Keele University, Staffordshire, ST5 5BG, UK}
\affil{$^{\dagger}$Corresponding author}
\affil{$^{\ddagger}$Royal Society University Research Fellow}
\email{\myemail}

\begin{abstract}
We present an analysis of the anomalous microlensing event, MOA-2010-BLG-073,
announced by the Microlensing Observations in Astrophysics survey on 2010-03-18.
This event was remarkable because the source was previously known to be
photometrically variable.  Analyzing the pre-event source lightcurve, we
demonstrate that it is an irregular variable over time scales $>$200\,d.  
Its dereddened color, \VISo, is \VISovale\,mag and
from our lens model we derive a source radius of \RSvale\,\Rsun,
suggesting that it is a red giant star. 
We initially explored a number of purely microlensing models for the event but found a
residual gradient in the data taken prior to and after the event.  This is
likely to be due to the variability of the source rather than part of the
lensing event, so we incorporated a slope parameter in our model in order to
derive the true parameters of the lensing system.  
We find that the lensing system has a mass ratio of \q=\qvale. The Einstein
crossing time of the event, \TE=\TEvale\,d, was sufficiently long that the
lightcurve exhibited parallax effects.  In addition, the source trajectory
relative to the large caustic structure allowed the orbital motion of the
lens system to be detected.  Combining the parallax with the Einstein
radius, we were able to derive the distance to the lens,
\DL=\DLvale\,kpc, and the masses of the lensing objects. The primary of the
lens is an M-dwarf with \MLp=\MLpvale\,\Msun\ while the companion has
\MLs=\MLsvale\,\MJ, putting it in the boundary zone between planets and 
brown dwarfs. 
\end{abstract}


\section{Introduction}


The mass function of individual compact objects (brown dwarfs and
planets) in the Galaxy remains poorly understood, particularly at the low-mass
end.  Brown dwarfs (BD) are commonly defined as objects with masses between the
deuterium and hydrogen burning limits (DBL and HBL, respectively) but these can
be hard to detect, being intrinsically faint and fading further as they cool
over time.  Below the DBL, the mass function for individual objects is even more
poorly measured.  Unbound, free-floating objects of planetary mass
have been discovered via direct imaging of clusters (for example
in $\sigma$~Orionis \citep{Bejar2012}) and in the field (e.g. the
6--25\,\MJ\ object reported by \citet{Kirkpatrick2006}).  \citet{Sumi2011}
reported a population of planets which are either unbound or at very wide
separations, discovered when their gravity caused short timescale
microlensing events.  

At least partially as a result of these poor constraints the origin of low-mass
compact objects remain unclear.  Although traditionally thought of as separate
classes of objects, planets and brown dwarfs form a continuous scale of mass and
are best distinguished by the circumstances of their formation
\citep{Burrows2001, Chabrier2005, Sahlmann2010, Chabrier2011}.  Planets form in disks of
material orbiting a protostellar object and may subsequently migrate to
different period orbits.  Brown dwarfs on the other hand, are considered to be
the extreme low mass end of the star formation process by fragmentation of
locally over dense cores caused by turbulence in a cloud \citep{Chabrier2011},
which can themselves form protoplanetary disks \citep{Klein2003, Scholz2006}. 
This mechanism may also produce objects of a few Jupiter masses.  
For a recent review, see \citet{Luhman2012}.  


The mass function of free-floating low-mass objects is likely to be different
from those bound to stars.  \citet{MarcyButler2000} identified a
paucity of BDs orbiting close ($<$3\,AU) to their host stars, a region where
Jovian-mass planets are commonly found.  This ``brown dwarf desert'' may
represent the gap between the largest objects that can form in protoplanetary
disks and the smallest objects that can concurrently collapse/condense next to
a star.


Two different theories have been proposed to explain the formation of Jovian
planets in disks (for a review, see \citet{Zhou2012}).  The core accretion model predicts planets form from
protoplanetary cores, growing up to tens of Jupiter masses
\citep[e.g.]{Mordasini2009, Baraffe2010} but predicts few giant planets around
M-dwarfs.  Higher mass stars are thought to have disks with enhanced surface
densities which allow the cores to grow more rapidly \citep{Laughlin2004}, as do
disks with a high fraction of dust, leading to enhanced planet formation around
high metallicity stars \citep{IdaLin2004}.  Alternatively, the model of planet
formation via gravitational instabilities in the disk \citep[e.g.]{Boss2006}
tends to favor the formation of more massive planets, in generally wider
orbits.  


A number of lines of evidence support the core accretion theory.  There is a
well-established correlation of increasing planet frequency with stellar 
metallicity \citep{Santos2001, FischerValenti2005, Maldonado2012}.  The results
of radial velocity surveys imply there is a derth of M-dwarf stars with massive,
close-in planets.  In part, this reflects an observational bias against these
faint objects but the sample is sufficiently large that a real statistical trend
is emerging \citep{Cumming2008, Johnson2010, Bonfils2011}, for companions with
$P<2000$\,d. Recent spectroscopic and {\em Kepler} results have confirmed the
prediction of a rapid increase in frequency for planets with small radii (down
to 2\Rearth) and $P<50$\,d for all spectral types, and found that these small
planets are several times more common around stars of late spectral type
\citep{Bonfils2011, Howard2012}.  

However, the core accretion model has difficulty forming massive planets at
large orbital radii and such systems have been discovered, for example HR~8799
\citep{Marois2008}.  Furthermore, a number of planets have been found orbiting
M-dwarf hosts at larger orbital separations, for example
\citet{Dong2006,Forveille2011,Batista2011}.   These systems may
instead form through gravitational instability in the disk, which can account
for companions up to several Jupiter masses around M-dwarfs \citep{Boss2006}. 

To better understand the formation mechanisms of heavy substellar companions in
bound systems we need to trace the distributions of the physical and orbital
properties (such as mass ratio, orbital separation, occurrence frequency) of a
significant number of systems.   Yet relatively few bound brown dwarf companions
have been reported, despite their being easy to detect at close orbital
separations (the ``brown dwarf desert'').  

Microlensing offers a complementary window onto BD and planet formation by
probing for cooler companions of all masses in orbital radii between $\sim$0.2
-- 10\,AU, separations which are difficult or time consuming to explore by other
methods \citep{Shin2012a}.  It can probe the companion mass function down to
M- and brown dwarf hosts, and is sensitive to companions from nearly equal mass
down to terrestrial masses.  

Sixteen systems have been published to date\footnote{Listed on
exoplanet.eu}, and thanks to large-scale galactic lensing surveys and efficient
follow-up, each season's Bulge observing campaign is now producing a regular
yield of new discoveries (e.g. \citet{Bachelet2012, Yee2012, Miyake2012}). Of
these 16 companions, 3 are giant planets orbiting M-dwarf stars:
OGLE-2005-BLG-071Lb, a 3.8\,\MJ\ planet \citep{Dong2009}, MOA-2009-BLG-0387Lb,
with $M_{P}$=2.5\,\MJ\ planet \citep{Batista2011} and MOA-2011-BLG-293Lb, which
hosts a 2.4\,\MJ\ companion \citep{Yee2012}.

Here we present the newly discovered system,
\objectname[MOA-2010-BLG-073L]{MOA-2010-BLG-073L}, an M-dwarf star with a companion
whose mass is close to the deuterium burning limit of $\sim$12.6\,\MJ.  The discovery
and follow-up observations are described in \textsection~\ref{sec:obs} and we discuss
the variability of the source star in \textsection~\ref{sec:variability}.  We
describe our analysis in \textsection~\ref{sec:analysis},
from which we derive the physical properties of the lens in
\textsection~\ref{sec:physicalparams}.  Finally, we discuss our findings in
\textsection~\ref{sec:discussion}.

\section{Observations} \protect\label{sec:obs}

The microlensing event, \objectname[MOA-2010-BLG-073]{MOA-2010-BLG-073}, was
first announced by the Microlensing Observations in
Astrophysics\footnote{www.phys.canterbury.ac.nz/moa} (MOA, \citet{Bond2001,
Sumi2003} on the 1.8\,m telescope at Mt. John  Observatory, New Zealand) survey
on 2010-03-18.  A background source star in the Galactic Bulge,
$\alpha$=18:10:11.342, $\delta$=-26:31:22.544 (J2000.0), previously having a
mean baseline magnitude of $I\sim$16.5\,mag, was discovered to be rising
smoothly in brightness consistent with a point-source, point-lens (PSPL)
microlensing event.  However, on 2010-05-03 the event was found to show an
anomalous brightening of $\sim$ 0.5\,mag and an alert was issued (K.~Furusawa,
private comm.).  \\

Microlensing follow-up teams worldwide --
RoboNet-II\footnote{robonet.lcogt.net} \citep{Tsapras2009},
$\mu$FUN\footnote{www.astronomy.ohio-state.edu/$\sim$microfun} \citep{Gould2006},
PLANET\footnote{planet.iap.fr} \citep{Beaulieu2006} and
MiNDSTEp\footnote{www.mindstep-science.org} \citep{Dominik2010}
-- responded to provide intensive coverage of the event for the duration of the
anomaly ($\sim$2\,days), and monitored the event as it returned to baseline,
over the course of the next $\sim$2\,months.  \\


In addition to the MOA data, taken with the wide band ``MOA-Red'' filter
(corresponding to $R$+$I$ bandpasses), the event was observed from several
other sites in New Zealand.  The 0.41\,m telescope at Auckland Observatory,
the 0.35\,m at Kumeu Observatory and the 0.41\,m Possum Observatory all 
used $R$-band filters while the 0.304\,m Molehill Astronomical Observatory
(MAO),   the 0.35\,m telescope at Farm~Cove Observatory (FCO) and the 0.4\,m
telescope at Vintage Lane Observatory (VLO) all observed it unfiltered.  
The event was then picked up from three sites in Australia, firstly in
$I$-band by the 1\,m Canopus telescope in Tasmania followed by  the 2\,m
Faulkes Telescope South (FTS), where an SDSS-$i$ filter was used.  The
0.6\,m telescope in Perth also observed in $I$ band.   Of the observing sites
around longitude zero, the event was imaged from the 1\,m telescope at the
South African Astronomical Observatory (SAAO) using an I band filter and in
$J$, $H$ and $K_{S}$  by the 1.4\,m Infra Red Survey Facility (IRSF), also
at SAAO.   The 2\,m Liverpool Telescope (LT) observed the event in SDSS-$i$
from the Canary Islands.  

\begin{figure*}
\figurenum{1}
\begin{center}
\includegraphics[width=14cm]{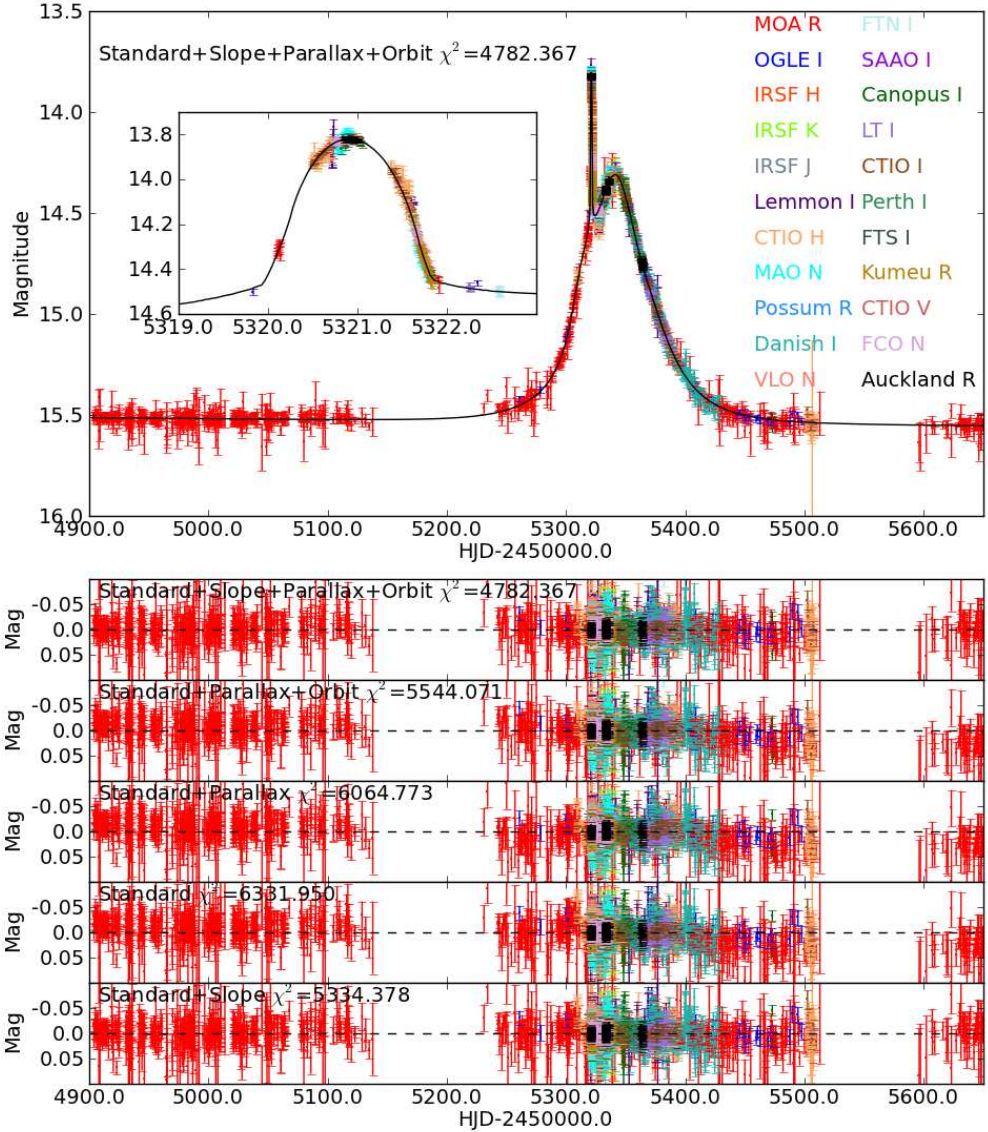}
\caption{Top panel: The complete light curve of MOA-2010-BLG-073 with the data sets from the various observatories overlaid with our best fitting model. 
The inset shows the anomaly in greater detail.  Lower panels: The residuals of the fit for each of the best-fitting
models of each class.}
\protect\label{fig:finallc}
\end{center}
\end{figure*}

As darkness fell in the Americas, a number of Chilean telescopes picked up the
observing baton: the SMARTS 1.3\,m at the Cerro Tololo Interamerican Observatory (CTIO)
obtained data in $V$, $I$ and $H$ bands with the {\sc andicam} camera, and the Danish
1.54\,m used an $I$ filter.   Though in the midst of commissioning the new OGLE-IV
camera at the time, the 1.3\,m Warsaw telescope also covered the 
event in $I$-band\footnote{ogle.astrouw.edu.pl}.  The 1\,m Mt.~Lemmon Telescope in Arizona
imaged the event in the $I$-band and in the extreme west, the 2\,m Faulkes
Telescope North (FTN), Hawai'i used an SDSS-i filter to complete the 24-hour
coverage of the event from the Pacific.  Table~\ref{tab:obstable} summarizes the
data obtained, which are plotted in Figure~\ref{fig:finallc}. 

The high density of Galactic Bulge star fields and the consequent degree of overlap
(or blending) in stellar point-spread functions (PSF) has long since made difference
image analysis (DIA) the photometry method of choice among microlensing teams.  Both
MOA and OGLE make their photometry available to the community, automatically reducing
their data with their custom pipelines described respectively in \citet{Bond2001} and
\citet{Udalski2003}.  The RoboNet data (from FTN, FTS and the LT) were reduced with
the project's automated data reduction pipeline, which is based around the
DanDIA package \citep{Bramich2008}.   This software was also later used to reduce
data from Canopus, the Perth 0.6\,m, the SAAO 1\,m and the $H$-band data from CTIO,
while the DIAPL package was used to process the images from the Danish telescope. 
The PLANET team released their photometry (produced by the WISIS pipeline) in real
time via their website, and the Pysis DIA pipeline \citep{Albrow2009} was used for
later re-reduction of these data sets. 

\begin{deluxetable}{lccccccc}
\tabletypesize{\small}
\tablewidth{13cm}
\tablenum{1}
\tablecolumns{8}
\tablecaption{Summary of observations.\label{tab:obstable} }
\tablehead{\colhead{Telescope \&} & \colhead{Filter} & \colhead{$u_{\lambda}^{(c)}$} & 
      	\colhead{$\Gamma_{\lambda}^{(c)}$} &
      	\colhead{N frames} & \colhead{N frames} & \colhead{$a_{0}^{(c)}$} &
	\colhead{$a_{1}^{(c)}$} \\
	\colhead{aperture [m]} & \colhead{} & \colhead{} & \colhead{} & \colhead{total} & \colhead{used} &
	\colhead{} & \colhead{[mag]} }
\startdata
MOA 1.8            	& $R$/$I$     & 0.7027  & 0.6118  & 1747  & 1726  & 1.305 & 0.005 \\
OGLE 1.3$^{a}$        	& $I$	      & 0.6098  & 0.5103  & 47    & 42    & 2.188 & 0.005 \\
Auckland 0.41 	      	& $R$	      & 0.7027  & 0.6118  & 136   & 136   & 0.910 & 0.000 \\
Canopus 1.0   	      	& $I$	      & 0.6098  & 0.5103  & 162   & 159   & 1.310 & 0.005 \\
CTIO 1.3      	      	& $V$	      & 0.7817  & 0.7048  & 19    & 18    & 0.603 & 0.000 \\
CTIO 1.3      	      	& $I$	      & 0.6098  & 0.5103  & 162   & 162   & 1.010 & 0.003 \\
CTIO 1.3      	      	& $H$	      & 0.4145  & 0.3206  & 586   & 575   & 1.340 & 0.016 \\
Danish 1.54   	      	& $I$	      & 0.6098  & 0.5103  & 498   & 491   & 1.130 & 0.014 \\
Farm Cove 0.4$^{b}$ 	& Unfiltered  &    -    & 0.5611  & 225   & 225   & 0.975 & 0.000 \\
FTN 2.0       	      	& SDSS-$i$    & 0.6098  & 0.5103  & 159   & 158   & 1.055 & 0.011 \\
FTS 2.0       	      	& SDSS-$i$    & 0.6098  & 0.5103  & 129   & 129   & 1.125 & 0.006 \\
IRSF 1.4      	      	& $J$	      & 0.4836  & 0.3844  & 4	  & 4	  & 1.000 & 0.000 \\
IRSF 1.4      	      	& $H$	      & 0.4145  & 0.3206  & 4	  & 4	  & 1.000 & 0.000 \\
IRSF 1.4      	      	& $K_{S}$     & 0.3550  & 0.2684  & 4	  & 4	  & 1.000 & 0.000 \\
Kumeu 0.35    	      	& $R$	      & 0.7027  & 0.6118  & 272   & 272   & 0.772 & 0.000 \\
Lemmon 1.0    	      	& $I$	      & 0.6098  & 0.5103  & 116   & 105   & 1.290 & 0.020 \\
LT 2.0	      	      	& SDSS-$i$    & 0.6098  & 0.5103  & 167   & 167   & 1.155 & 0.006 \\
MAO 0.304$^{b}$     	& Unfiltered  &    -    & 0.5611  & 238   & 238   & 1.025 & 0.018 \\
Perth 0.6     	      	& $I$	      & 0.6098  & 0.5103  & 66    & 66    & 1.095 & 0.008 \\
Possum        	      	& $R$	      & 0.7027  & 0.6118  & 15    & 15    & 1.030 & 0.009 \\
SAAO 1.0      	      	& $I$	      & 0.6098  & 0.5103  & 30    & 30    & 1.050 & 0.011 \\
Vintage Lane 0.4$^{b}$	& Unfiltered  &    -    & 0.5611  & 124   & 124   & 0.995 & 0.000 \\
\enddata
\tablenotetext{a}{Includes only OGLE-IV data taken during event MOA-2010-BLG-073.}
\tablenotetext{b}{For unfiltered or very broadband observations we adopted a limb-darkening 
parameter value which was the average of that for R and I bands: $(\Gamma_{R}+\Gamma_{I})/2$.}
\tablenotetext{c}{$u_{\lambda}$, $\Gamma_{\lambda}$ are defined by Eqns.~\ref{eqn:LD1}, 
\ref{eqn:LD2}, \ref{eqn:LDC} and $a_{0}$, $a_{1}$ by Eqn.~\ref{eqn:photerrors}.}
\end{deluxetable}

\section{Variability of the Source Star}
\protect\label{sec:variability}

\begin{figure*}
\figurenum{2}
\epsscale{1.0}
\plotone{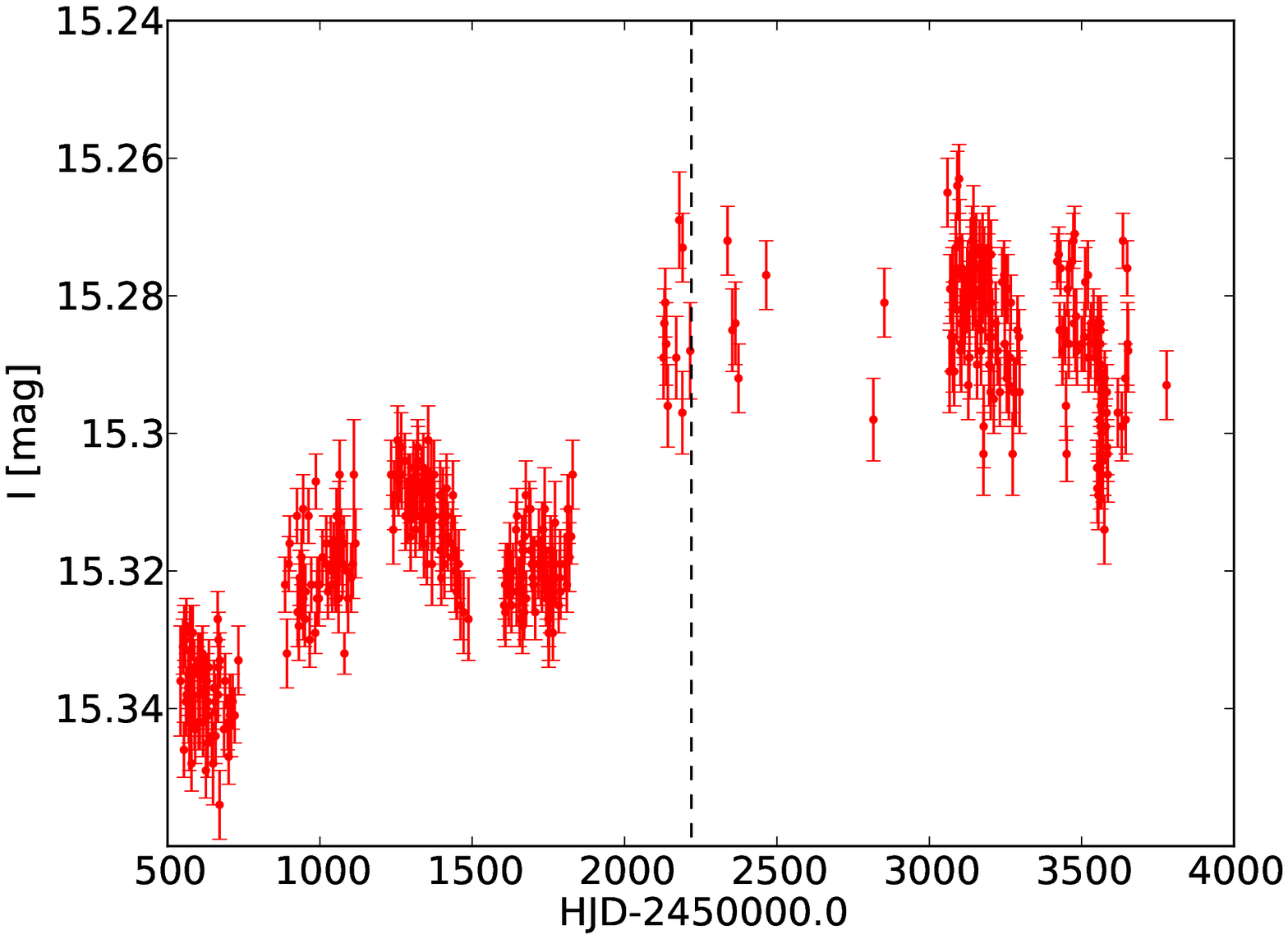}
\caption{All available OGLE-II (1998--2001) and OGLE-III (June 2001--2006) 
lightcurve data for the source star of MOA-2010-BLG-073, taken prior to the 
event.  The instrument upgrade to OGLE-III occurred around HJD=2452000.0.  
The apparent offset in magnitude at this time may be due in part to the 
difficulties of accurately calibrating photometry between the two wide field
surveys.  \protect\label{fig:OGLEbaseline}  }
\end{figure*}

MOA-2010-BLG-073 was present in the fields of the OGLE-II and OGLE-III surveys
so the source star's I-band photometric record extends from 1998 to 2006
(Fig.~\ref{fig:OGLEbaseline}).  OGLE-IV was in the commissioning phase when this
event took place. From this excellent baseline it was immediately clear that
the source is variable over many-month time scales.  This raised the possibility
that shorter-term variability might obfuscate the microlensing signal, making it
difficult to determine its properties.  

\begin{figure*}
\figurenum{3}
\epsscale{1.0}
\plotone{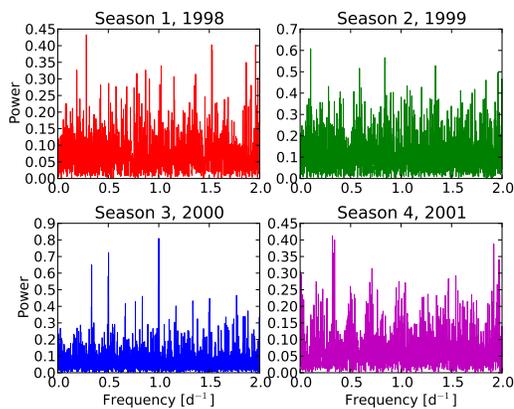}
\caption{Periodograms of the OGLE-II (1998--2001) seasons of data, analyzed separately with
the ANOVA algorithm.  No significant periodicities were found in the range
$P$=0.5--200\,d, aside from the expected aliases. \protect\label{fig:seasonalperiod}  }
\end{figure*}

To investigate this possibility, we performed a search for periodicities in the
baseline OGLE-II and OGLE-III data, excluding the lensing event, using the ANOVA
algorithm \citep{SC1996}.  Due to the seasonal gaps in the baseline, we
analyzed  the OGLE-II data in yearly subsets as it is the best sampled,
searching for periods between $P$=0.5--200\,d. As
Figure~\ref{fig:seasonalperiod} demonstrates, there are no significant or
persistent periodicities, other than the expected integer multiples of the 1-day
sampling alias.  

We then combined the OGLE-II and III data sets in order to search for
periods up to $P$=4000\,d.   Figure~\ref{fig:OGLEbaseline} indicates a
slight ($\sim$0.04\,mag) magnitude offset between the OGLE-II and III data. 
This can occur as a residual of OGLE's photometric calibration between the
two surveys but it might also be the result of the intrinsic stellar
variation.  Therefore, we performed a search for periods between 0.5 and
4000\,d based on the combined data both with and without this offset
(estimated visually).  In both cases the periodogram was dominated by the
window function; the only significant power was found in the peaks marking
multiples of the 1-day alias, plus one peak at extreme low frequency
corresponding to the finite length of the data set.  We conclude that this
star is an irregular, long-term variable, most likely as a result of pulsations.  

\begin{figure}
\figurenum{4}
\epsscale{1.0}
\plotone{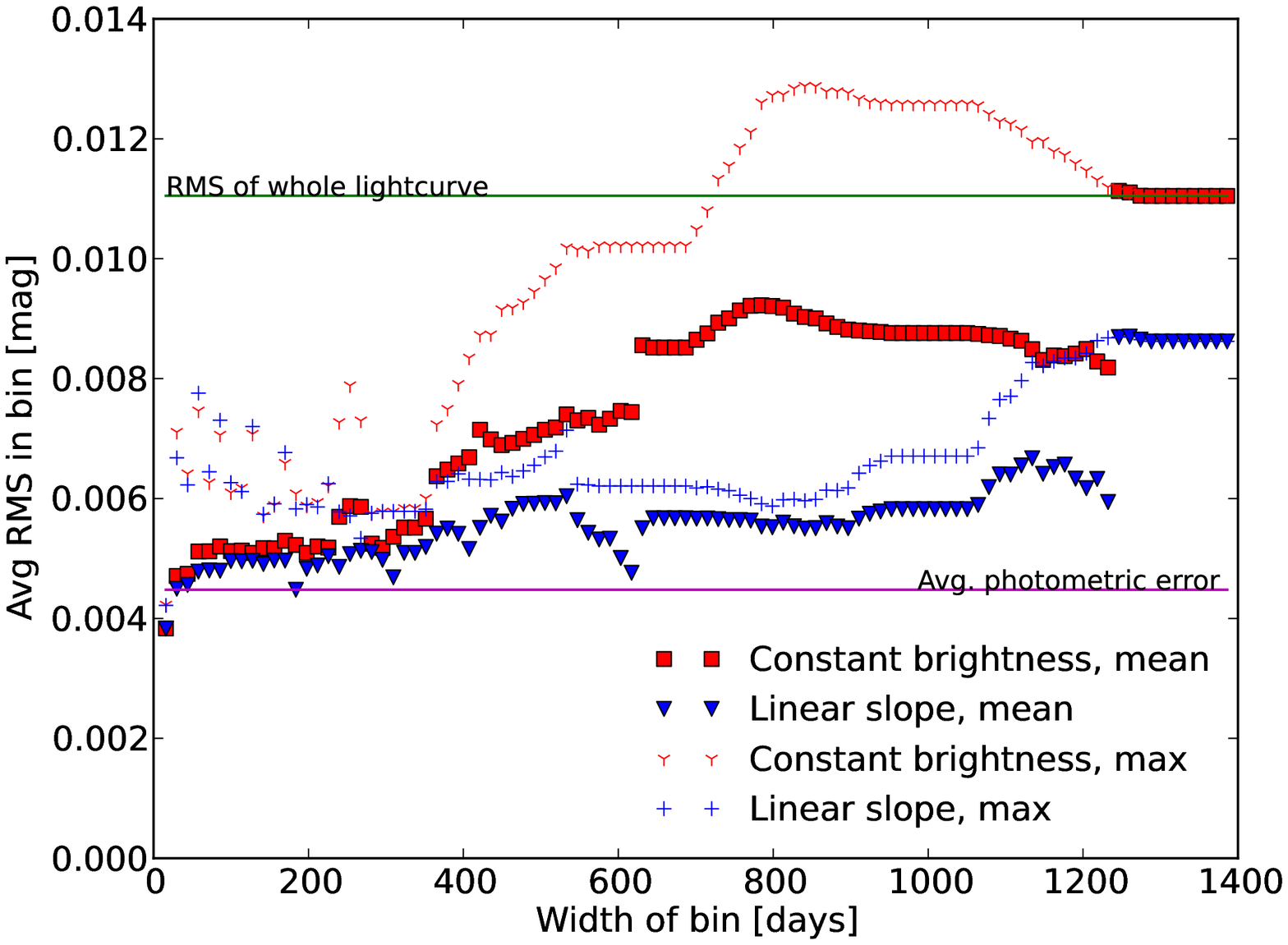}
\caption{The OGLE-II photometry was binned using a series of bins of varying
width in time.  The data in each bin was fitted with two functions a) constant
brightness b) with a linear slope, and the RMS around these curves was averaged
over all the bins in the lightcurve.  The average and maximum RMS for each binning is 
plotted here against the width of the bin, indicating variability over different 
timescales. 
 \protect\label{fig:RMSvstime}}
\end{figure}

However, there remained the possibility that the star could be irregularly
variable on time scales comparable to that of the lensing event.  To test
this possibility, we binned the OGLE-II light curve on a range of time
scales between 2--1200\,d.  To each bin we fitted two functions: one of
constant brightness, and one with a linear slope and calculated the weighted
root-mean-square (RMS) photometric scatter around each function per bin.  We
plot the average and maximum RMS (calculated over all bins in the
lightcurve) against the width of the bins in time in
Figure~\ref{fig:RMSvstime}.    On time scales shorter than 200\,d the RMS
scatter in the binned light curve is reasonably constant  implying no
significant short term variability.  The longer term trend becomes clearly
evident in the constant brightness curves for time scales longer than
400\,d. We note that for bin widths between $\sim$800--1200\,d, this curve
has an RMS actually exceeding that of the whole light curve; this is because
these bins were sufficiently wide that the first bin included the majority
of the data, and the most variable sections of the light curve.  As the bin
widths became longer, they included more data points from the relatively
stable section towards the end of the OGLE-II data set, and the RMS drops. 
The deviation of the OGLE lightcurve from a constant brightness exceeds
3$\sigma$ for timescales longer than $\sim$750\,d.  However, it deviates
from a linear slope by $\leq$1.6$\sigma$ for timescales less than 1000\,d,
so we represent this variation as a gradient in the lightcurve over the
duration of the event.  

The most intuitive way to account for the variation of the source was to
measure the gradient of the lightcurve taken at baseline immediately before and
after the event.  Unfortuantely only one of the available datasets covered
these periods.  Fitting a straight-line model (via a non-linear
least-squares Marquardt-Levenberg algorithm) to the MOA 2009 and 2011 season
data, we measured a slope of 0.018\,mag/yr.  However, the RMS scatter in the
residuals of this fit were 0.016\,mag, making it difficult to properly
determine the slope.  Additionally, in order to remove this slope from the
other datasets, it would be necessary to transform the fluxes measured by
each telescope on to the same scale as the MOA data.  We attempted this via
a linear regression approach but found that significant residuals remained. 
These contributed to overall higher \chisq\ values when the corrected data
were fit with binary lensing models.  We therefore adopted the alternative
method of incorporating the slope as an additional parameter in our lensing
model which we fit to the original, uncorrected data and we describe this
approach in the following sections. 

\section{Analysis}
\protect\label{sec:analysis}

In the analysis of this event, we used the established modeling software
developed by S.~Dong and C.~Han \citep{Dong2006, Shin2012b}.

\subsection{Initial Parameters}

As a starting point for our analysis, we needed approximate values for the
three parameters of the standard model for a PSPL event (not yet including
the slope; this is discussed in \textsection~\ref{sec:slope}): \To, the time
of peak magnification occurring at the closest projected separation between
the lens and source, \uo\ and the Einstein radius crossing time, \TE.  
Following standard convention, all distances are quoted in units of the
angular Einstein radius, \thetaE, of the lens.  

To estimate these parameters, we combined all available data sets into a
single light curve using the following scaling to take account of the varying
degrees of PSF blending from different instruments:

\begin{equation}
\protect\label{eqn:blending}
f(t,k) = A(t)f_{s}(k) + f_{b}(k), 
\end{equation}

\noindent where $f(t,k)$ is the measured flux of the target at time $t$
from data set $k$, $A(t)$ is the lensing magnification at that time,
\fluxs$(k)$ is the flux of the source star and \fluxb$(k)$ represents the flux
of all stars blended with the source in the data set.  A regression fit was used
to measure \fluxs\ and \fluxb\ for each data set, producing an aligned light
curve.   Although the resulting parameter estimates are somewhat different from
their `true' values due to the existence of the anomalous deviation on the light
curve, they provided a starting point in parameter space. 

We note that two additional parameters can contribute to a PSPL model. They are
the lens parallax parameters \pien\ and \piee\ that account for the light curve
deviation caused by the motion of the Earth in its orbit over the course of the
event.  The vector microlens parallax, $\vec{\pi_{\rm{E}}} =
AU/\widetilde{r_{\rm{E}}}$, where  $\widetilde{r_{\rm{E}}}$ is the Einstein radius
projected onto the observer plane and:

\begin{equation}
\protect\label{eqn:parallax}
\vec{\pi}_{\rm E} \equiv (\pi_{{\rm E},N}, \pi_{{\rm E},E}) \equiv (\cos{\phi_{\pi}}, \sin{\phi_{\pi}})\pi_{\rm{E}}, 
\end{equation}

where $\phi_{\pi}$ represents the direction of lens motion relative to the
source as a counter-clockwise angle, north through east.  However, this is
generally significant only for long ($\sim$months) time scale events, and was
included at a later stage (see Section~\ref{sec:parallax}).  

\subsection{Finite Source}

The sharp spike feature in the lightcurve is indicative of the source
closely approaching  or crossing a caustic.  In these circumstances, it
cannot be approximated by a point light source and must be treated as a disk
of finite angular radius, $\rho$, with wavelength-dependent limb-darkening. 
This is addressed within our software using the ray-shooting approach
\citep{Kayser1986}: the path of light rays is traced from the image plane
back to the source, taking into account the bending of the trajectory
according to the lens equation.  If a ray is found to ``land'' within the
radius of the source, its intensity is computed taking limb-darkening
into account.  We derived this from the linear limb darkening law:

\begin{equation}
\protect\label{eqn:LD1}
I_{\lambda}(\cos{\phi}) = I_{\lambda}(1)\left[1 -
u_{\lambda}(1-\cos{\phi})\right],
\end{equation}

where $I_{\lambda}$ is the intensity of the source at radius $\phi$ from the
center, relative to the central intensity $I_{\lambda}(1)$ in the same wavelength,
$\lambda$, scaled by the coefficient $u_{\lambda}$.  While more accurate limb
darkening models are available, they are not commonly used in microlensing
analyses due to the complexity introduced by combining data from many sources
(\citet{Bachelet2012} discussed this in more detail).  The values of $u_{\lambda}$
for each passband were calculated from the Kurucz ATLAS9 stellar atmosphere models
presented by \citet{Kurucz1979} using the method of \citet{Heyrovsky2007}.  However,
within the microlensing community and software, Equation~\ref{eqn:LD1} more
commonly follows the formalism derived by \citet{Albrow1999}:

\begin{equation}
\protect\label{eqn:LD2}
I_{\lambda} = \frac{F_{\lambda}}{\pi\theta_{*}^{2}}
\left[1-\Gamma_{\lambda}\left(1-\frac{3}{2}\cos{\phi}\right)\right],
\end{equation}

where $F_{\lambda}$ is the total flux from the source in a given passband and
$\phi$ is the angle between the line of sight to the observer and the normal to
the stellar surface.  The limb darkening coefficient, $\Gamma_{\lambda}$ is
related to $u_{\lambda}$ by:

\begin{equation}
\protect\label{eqn:LDC}
\Gamma_{\lambda} = \frac{2u_{\lambda}}{3-u_{\lambda}}.
\end{equation}

The values of $u_{\lambda}$ and $\Gamma_{\lambda}$ applied for each dataset are
presented in Table~\ref{tab:obstable}.  The lensing magnification is then computed
as the ratio of the number of rays reaching the source plane relative to the
number in the image plane.  This approach is only required while the source is
close to the caustic.  At larger separations, the software employs a semi-analytic
hexadecapole approximation to the finite source calculation to improve computation
speeds \citep{Pejcha2009, Gould2008}. 

\subsection{Standard Binary Model Grid Search}

To model the light curve of a binary lens event, we introduced three additional
parameters: $q=M_{L,2}/M_{L,1}$, the ratio of the masses of the two bodies
composing the lens where \MLp\ is the more massive component, \s, the projected
separation of those masses and \alf, the angle of the trajectory of the lensed
source star, relative to the lens' binary axis.  The  frame of reference was
defined to be at rest with respect to the Earth at time \topar, which we took to
be the time of caustic crossing at HJD=2455321.0, estimated from the easily
identifiable feature in the light curve (following the notation of
\citealt{Skowron2011}).  

With seven variables in the model (\To, \uo, \TE, \alf, \s, \q, $\rho$), a
number of different lens/source configurations may produce similar light curves,
so it was necessary to thoroughly explore a large area of parameter space in
order to ensure all possible solutions are identified.  We therefore constructed
a grid of models, spanning set ranges in the values of the three variables upon
which the overall \chisq\ of the fit depended most sensitively, \s, \q\ and
\alf.  Each node in this grid took fixed values of (\s,\q,\alf) and used a
Markov Chain Monte Carlo (MCMC) approach \citep{Dong2006} to find the best
fitting model by optimizing the other parameters.  To improve efficiency, a
magnification map is generated by rayshooting for each point in the grid from
which the model light curves used to compute the \chisq\ are drawn.  The grid
covered the following range: $\log(s_{0}) = -0.6:0.6$ in steps of 0.012,
$\log(q) = -4.0:1.0$ in steps of 0.05 and $\alpha = 0.0:6.3$ in steps of 0.6.

Mapping out the \chisq\ for each node in this grid, we found a number of
local minima.  Visual inspection of these models overlaid on the light curve
demonstrated that some more closely followed the data than others.  Our
first pass analysis included substantial baseline photometry before and
after the event.  This was not well fit by the models due to the variability
of the source and hence the \chisq\ map gave a distorted view of regions in
parameter space that best match the event.  For this reason, we proceeded by
repeating the grid search using just data taken during 2010.  This produced
two clear minima in \chisq\ of which one model stood out as by far the best
match to the data. We then conducted a refined grid search over this
restricted region of parameter space, taking smaller incremental steps.  

\subsection{Optimized Standard Binary Model}

The refined grid search produced a reasonable model, fitting the majority of
the data from all telescopes.  This was used as a guide to identifying
likely outlying data points, for which the quality of the reduction was then
checked. A handful of data points were removed at this stage.  However, this
model included only fixed values for \s, \q\ and \alf.  To properly
determine the standard binary model for this event, our next step was to
allow the seven parameters (\To, \uo, \TE, \alf, \s, \q, $\rho$) to be
optimized during the MCMC fitting process, which used the grid search
results as its starting point.  At this point we included the extended
baseline data from MOA for seasons 2009 and 2011, as these fall within the
period for which the source's variability can be approximated with a
straight line; we address this in Section~\ref{sec:slope}.  

\subsection{Normalization of Photometric Errors}

When fitting microlensing events, the reduced \chisq\ of the fit on a per
data set basis, \redchisq, typically produces a range of values both less
than and exceeding the expected unity value.  This can occur as different
groups have slightly different ways of estimating photometric errors, but
can lead to over- or under-emphasis being placed on particular data sets
during the modeling process.  \\

A common technique to address this issue is to arrive at a complete model
for the event and then use this model to renormalize the original
photometric errors of each data set, $e_{\rm{orig}}$, according to the
expression:

\begin{equation}
\protect\label{eqn:photerrors}
e_{\rm{new}} = a_{0}\sqrt{e_{\rm{orig}}^{2} + a_{1}^{2}}.
\end{equation}

\noindent We first conducted the sequence of models described in the
following sections in order to find the best model for the event.  We then
set the coefficients $a_{0}$, $a_{1}$ such that the \redchisq\ relative to
that model equaled unity; the adopted values are given in
Table~\ref{tab:obstable}.  We then repeated our MCMC fitting process,
starting with the Standard binary model and systematically adding parameters
in to determine the extent of improvement in the model \chisq\ in each
case.  We compare models in Tables~\ref{tab:chisqtable} \&
\ref{tab:measuredpars}, and we plot the residuals ($\rm{data}-\rm{model}$)
in Figure~\ref{fig:finallc}.  In the following sections, the \chisq\ values
given are those post-renormalization.  \\

\subsection{Parallax}
\protect\label{sec:parallax}

Given that the event's \TE$\sim$44\,d$\sim$0.12\,yr, it was necessary to include
parallax in our model.  Using the parameters of the Standard Binary Model as a
starting point, we allowed our fitting process to optimize for \piee, \pien\
also.  We found that this significantly reduced the \chisq\ of the overall fit to
\chisqSPpos.   By default, this procedure explored models with positive projected
separations at closest approach of the lens and source, that is \uopos, which we
define as the source's trajectory passing the caustic at positive values of
$\theta_y$ in the lens plane (see Figure~\ref{fig:caustic}).  For the standard
model case the symmetry with respect to the binary axis of the caustic means that
the \uoneg\ solutions are identical.  However once parallax was included, this was
no longer the case, so we also explored \uoneg\ solutions (this degeneracy is
further discussed in \citet{Park2004}).  The parameters of the \uopos\ model were
taken as a starting point for the fit, except that the sign of \uo\ was reversed
and the \alf\ value became $2\pi-\alpha_{0}$.  We found the best fitting \uoneg\
model to be slightly less favored, with \chisq=\chisqSPneg.  

\subsection{Lens Orbital Motion}

The mass ratio and projected separation determined from this model put this
event close to the boundary between close and intermediate/resonant caustic
structure.  In this regime, small changes in the projected separation of the
lensing bodies due to their orbital motion can effectively change the shape of
the caustic (see Fig.~\ref{fig:caustic}) while the event is underway, sometimes
causing detectable deviations in the lightcurve.  To explore this possibility,
we included additional parameters in our model to describe the change in
projected binary separation, \dsdt\ and the rate of change of the angle of the
projected binary axis, \dadt.  Again we found that the \uopos\ model gave the
best fit, with \chisq=\chisqSPOpos, compared with \chisqSPOneg\ when \uoneg.  
This type of orbital motion is classified as ``separational'' in the schema put
forward by \citet{Gaudi2009, Penny2010}, and is detected in this event as the
source happens to cross the cusp of the caustic in the position where the
caustic changes most rapidly.  

\subsection{Sloping Baseline}
\protect\label{sec:slope}

While taking these second-order effects into account significantly improved the
fit to the data, the overall \chisq\ remained rather high.  Visual inspection of
the light curve still showed a gradient, especially in the 2010 baseline before,
relative to after, the event.  Based on our analysis in
\textsection~\ref{sec:variability}, this trend is likely to be part of the
longer-term variability of the source and not associated with the microlens.  In
order to determine the true lens/source characteristics though, this trend had
to be taken into account.  

The OGLE-II and III data demonstrate that the source variation over the
$\sim$150\,d time scale of the event can be approximated by a straight-line,
rather than a higher-order function.  We therefore introduced a `slope'
parameter to our model, representing the linear rate of change in magnitude
during the event.  This further improved the \chisq, and the best-fitting model
was once again the \uopos\ solution with \chisq=\chisqSPOSpos.  

We note that there exist degeneracies between the slope parameter and those
for parallax and orbital motion as they can be used to fit similar residuals in
the lightcurve.  To test for possible degeneracies, we also fit a standard model
plus the slope parameter alone and found that \chisq=\chisqSS.  The value for
the slope from this model, \SSlopevale, was consistent with that derived from
the model including parallax and orbital motion, \Slopevale.  

\subsection{Second Order Effects}

With the slope parameter included we had accounted for all the physical effects
which we expected to be present in the lightcurve. Having found that the
residuals showed no further variation at a level detectable above the
photometric noise, we did not attempt to include second order effects such as
xallarap etc.  

\subsection{Final Model}

\begin{figure*}
\figurenum{5}
\begin{center}
\includegraphics[width=14cm]{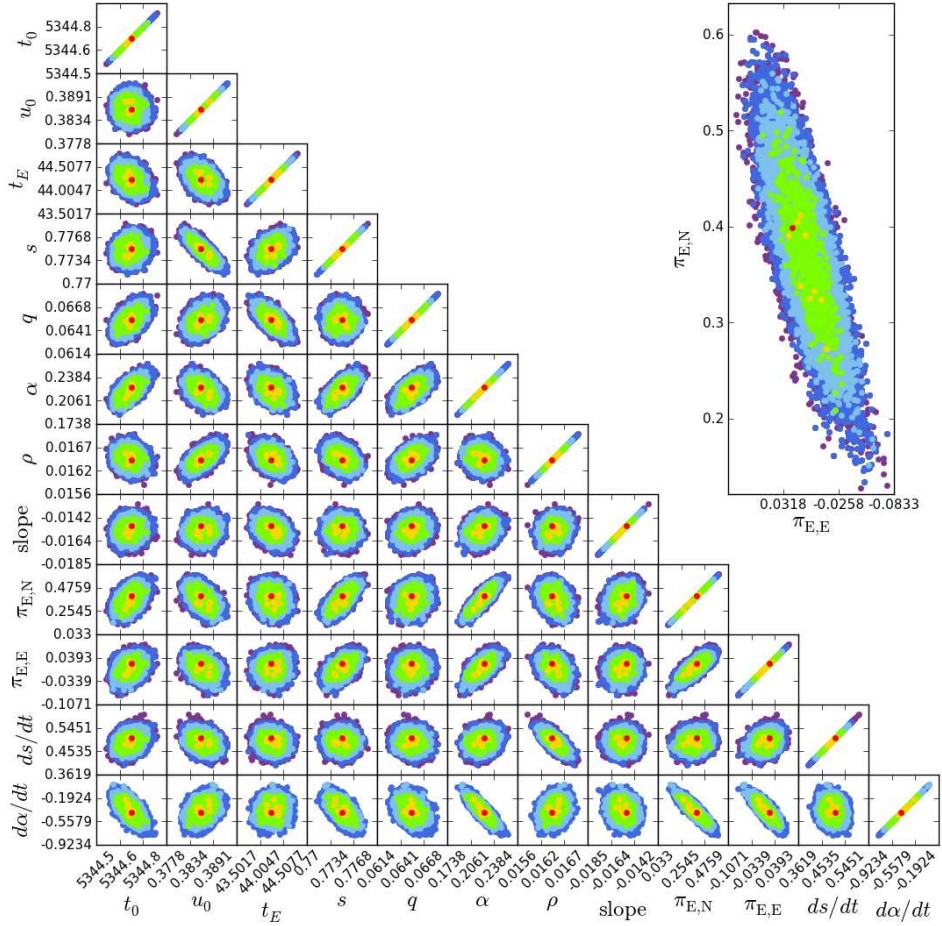}
\caption{Chi squared contours plotted as a function of the parameters fitted in the 
MCMC fit for the best model.  The red, orange, green, light blue, dark blue and purple 
colors indicate the regions with \dchisq$<$1--6$\sigma$ (respectively) from the best-fit solution. 
Inset: Close-up of the coutours for the parallax parameters.}
\protect\label{fig:paramgrid}
\end{center}
\end{figure*}

\begin{figure}
\figurenum{6}
\epsscale{1.0}
\plotone{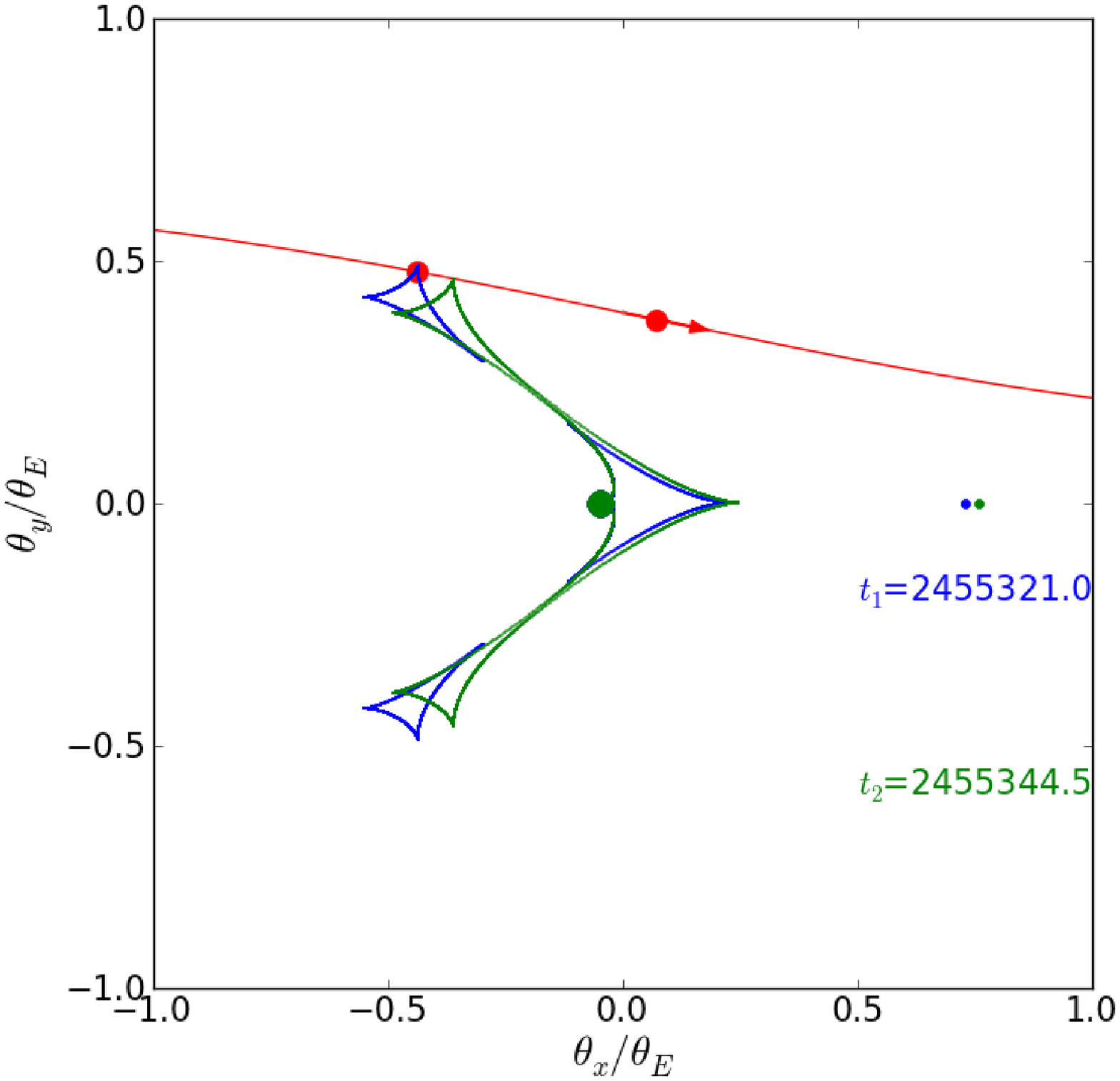}
\caption{Maps of the caustic structure during the anomaly and at the time of closest approach.  
The red line and arrow indicate the trajectory of the source in a reference frame centered on the
barycenter of the lensing system, while the red dots indicate the position of the source at these 
times. The green and blue dots indicate the positions of $M_{L,1}$
(largest) and $M_{L,2}$ at both times (radii of dots not to scale).  \protect\label{fig:caustic}  }
\end{figure}

All our models are compared in Table~\ref{tab:chisqtable} and the parameters
of the best-fitting models are presented in Table~\ref{tab:measuredpars} and
Figure~\ref{fig:finallc}.  We plot the \chisq\ for each link in the MCMC
chain for all parameters against one another in the best-fitting model in
Figure~\ref{fig:paramgrid}.  This plot was used as a diagnostic throughtout the
fitting process, as any correlations between parameters display distinct
trends as the chain moves towards the minimum.  The caustic structure
changed during the course of event, so in  Figure~\ref{fig:caustic} we show
the structure at two distinct times; the first at the time of the first
caustic crossing during the anomaly, and the second at the time of closest
approach.  

\begin{deluxetable}{lcc}
\tabletypesize{\small}
\tablewidth{11.5cm}
\tablenum{2}
\tablecolumns{3}
\tablecaption{The \chisq for the best-fitting model in each class, comparing \uopos\ and \uoneg\ solutions. \label{tab:chisqtable} }
\tablehead{\colhead{Model} & \multicolumn{2}{c}{\chisq} \\\colhead{} & \colhead{\uopos} & \colhead{\uoneg}}
\startdata
Standard      	      	      	      	      & \chisqS       & \chisqS \\
Standard+Parallax     	      	      	      & \chisqSPpos   & \chisqSPneg  \\
Standard+Parallax+Orbital Motion  	      & \chisqSPOpos  & \chisqSPOneg\\   
Standard+Parallax+Orbital Motion with Slope   & \chisqSPOSpos & \chisqSPOSneg \\  
Standard+Slope	      	      	      	      & \chisqSS      & \chisqSS \\
\enddata
\end{deluxetable}

\begin{deluxetable}{lccccc}
\tabletypesize{\scriptsize}
\tablewidth{13cm}
\tablenum{3}
\tablecolumns{6}
\tablecaption{The best fitting parameters. \label{tab:measuredpars}}
\tablehead{\colhead{Parameter} & \colhead{Standard} & \colhead{Standard} & \colhead{Standard} & \colhead{Standard+Parallax} & \colhead{Standard+Parallax} \\
\colhead{(Units)} & \colhead{} & \colhead{+Slope}& \colhead{+Parallax} & \colhead{+Orbital Motion} & \colhead{+Orbital Motion+Slope}}
\startdata
\chisq    	& 6331.950	& 5334.378    	  & 6064.773  	      	  & 5544.071		& 4782.367    \\
\dchisq$^{b}$   & \dchisqS    	& \dchisqSS   	  & \dchisqSP 	      	  & \dchisqSPO	      	& \dchisqSPOS  \\
\To\ (\hjdp)	& 5344.32     	& 5344.38     	  & 5344.47		  & 5344.83		& 5344.69     \\
	      	& $\pm$0.01   	& $\pm$0.01   	  & $\pm$0.02 	      	  & $\pm$0.03		& $\pm$0.02   \\
\uo\  	  	& 0.4089        & 0.403       	  & 0.403		  & 0.381	       & 0.386       \\
      	      	& $\pm$0.0009   & $\pm$0.001  	  & $\pm$0.001	      	  & $\pm$0.001         & $\pm$0.001  \\
\TE\ (d)  	& 43.82         & 44.84       	  & 43.49		  & 43.4		& 44.3        \\
      	      	& $\pm$0.08     & $\pm$0.09   	  & $\pm$0.09 	      	  & $\pm$0.1		& $\pm$0.1    \\
\s\   		& 0.7692        & 0.7725      	  & 0.7717		  & 0.7792		& 0.7750  \\
      	      	& $\pm$0.0005   & $\pm$0.0005     & $\pm$0.0006	      	  & $\pm$0.0007 	& $\pm$0.0007 \\
\q	        & 0.0705        & 0.0677      	  & 0.0695		  & 0.0683		& 0.0654  \\
      	      	& $\pm$0.0005   & $\pm$0.0005     & $\pm$0.0006	      	  & $\pm$0.0006 	& $\pm$0.0006 \\
\alf\ 		& 0.180         & 0.171       	  & 0.198		  & 0.297		& 0.221  \\
      	      	& $\pm$0.003    & $\pm$0.003  	  & $\pm$0.003	      	  & $\pm$0.006  	& $\pm$ 0.007 \\
$\rho$\     	& 0.01963     	& 0.01912     	  & 0.01931		  & 0.0163		& 0.0165 \\
      	      	& $\pm$0.00006	& $\pm$0.00007    & $\pm$0.00008	  & $\pm$0.0001 	& $\pm$0.0001 \\
\pien\    	&		&                 & 0.18		  & 0.96		& 0.37  \\
      	      	&     	      	&                 & $\pm$0.02		  & $\pm$0.04		& $\pm$0.05   \\
\piee\          &		&                 & -0.124		  & 0.09		& 0.01   \\
      	      	&     	      	&                 & $\pm$0.007  	  & $\pm$0.01		& $\pm$0.01   \\
\dsdt\ ($\rm{yr}^{-1}$)&	&                 &			  & 0.53		& 0.49    \\
      	      	&     	      	&                 &			  & $\pm$0.02		& $\pm$0.02   \\
\dadt\ ($\rm{yr}^{-1}$)&	&                 &			  & -1.21		& -0.37      \\
      	      	&     	      	&                 &			  & $\pm$0.06		& $\pm$0.08  \\
Slope\ (mag/yr)  &		& -0.0160     	  &			  &			& -0.0153 \\
      	      	&     	      	& $\pm$0.0004     &			  &			& $\pm$0.0004 \\
\enddata
\tablenotetext{a}{All timestamps are abbreviated to \hjdp = HJD-2450000.0.}
\tablenotetext{b}{Improvement in \chisq\ relative to that of the best fitting model.}
\end{deluxetable}

\section{Physical Parameters}
\protect\label{sec:physicalparams}

The purpose of this model is to ultimately arrive at the physical parameters of
the lens and source, which can be achieved using the known relations between
these and the lensing parameters obtained from the modeling.  \\

Chiefly of interest is the mass of the lensing system, \MLtot, which can be
determined explicitly for events where parallax is measurable provided the
angular extent of the Einstein radius, \thetaE, is known from:

\begin{equation}
M_{L,\rm{tot}} = \frac{c^{2}}{4G}\widetilde{r}_{\rm{E}}\theta_{\rm{E}} =
\frac{c^{2}AU}{4G}\frac{\theta_{\rm{E}}}{\pi_{\rm{E}}},
\end{equation}

\noindent where $\widetilde{r}_{\rm{E}}$ is the Einstein radius projected from
the source onto the observer's plane.  The model parameter $\rho$ represents the
angular size of the source star \thetaS\ in units of the angular Einstein radius
\thetaE.  We derive this from the crossing time taken for the source to travel
behind the lens, $t_{*}=\mu\theta_{S}$ where $\mu$ is the relative
source-lens proper motion.  $\rho$ can then be written as:

\begin{equation}
\rho = \frac{t_{*}}{t_{\rm{E}}} = \frac{\theta_{S}}{\theta_{\rm{E}}},
\end{equation}

\noindent These parameters also yield the distance to the lens, 

\begin{equation}
\frac{AU}{D_{L}} \equiv \pi_{L} = \theta_{\rm{E}}\pi_{\rm{E}} + \frac{AU}{D_{S}}
\end{equation}

\noindent which in turn yields the projected separation between the lens
components:

\begin{equation}
a_{\perp} = s_{0}D_{L}\theta_{\rm{E}}, 
\end{equation}

\noindent and the relative proper motion between lens and source, when
combined with \TE:

\begin{equation}
\mu=\frac{\theta_{\rm{E}}}{t_{\rm{E}}}.
\end{equation}

\noindent The appreciable lens orbital motion during this event also allows us to test
whether the companion object is bound to the primary lensing mass, via the
ratio of its kinetic to potential energy:

\begin{equation}
\rm{KE}/\rm{PE} = \frac{(s_{0}R_{E})^{3}\gamma^{2}}{8\pi^{2}M_{L}}, 
\end{equation}

\noindent where $\gamma$ relates the two lens orbital parameters:
$\gamma^{2}$=(\dsdt/\s)$^2$ + (\dadt)$^2$ and where the masses are in units of
\Msun, the distances in AU and time measured in years.  

\noindent However, these expressions include two key terms which are as yet,
unknown: \thetaS, the angular source radius and \DS, the distance to
the source. In order to extract the physical characteristics of the lens, we
therefore turned our attention to the characteristics of the source.  

\subsection{Source Star}

\begin{figure}
\figurenum{7}
\epsscale{1.0}
\plotone{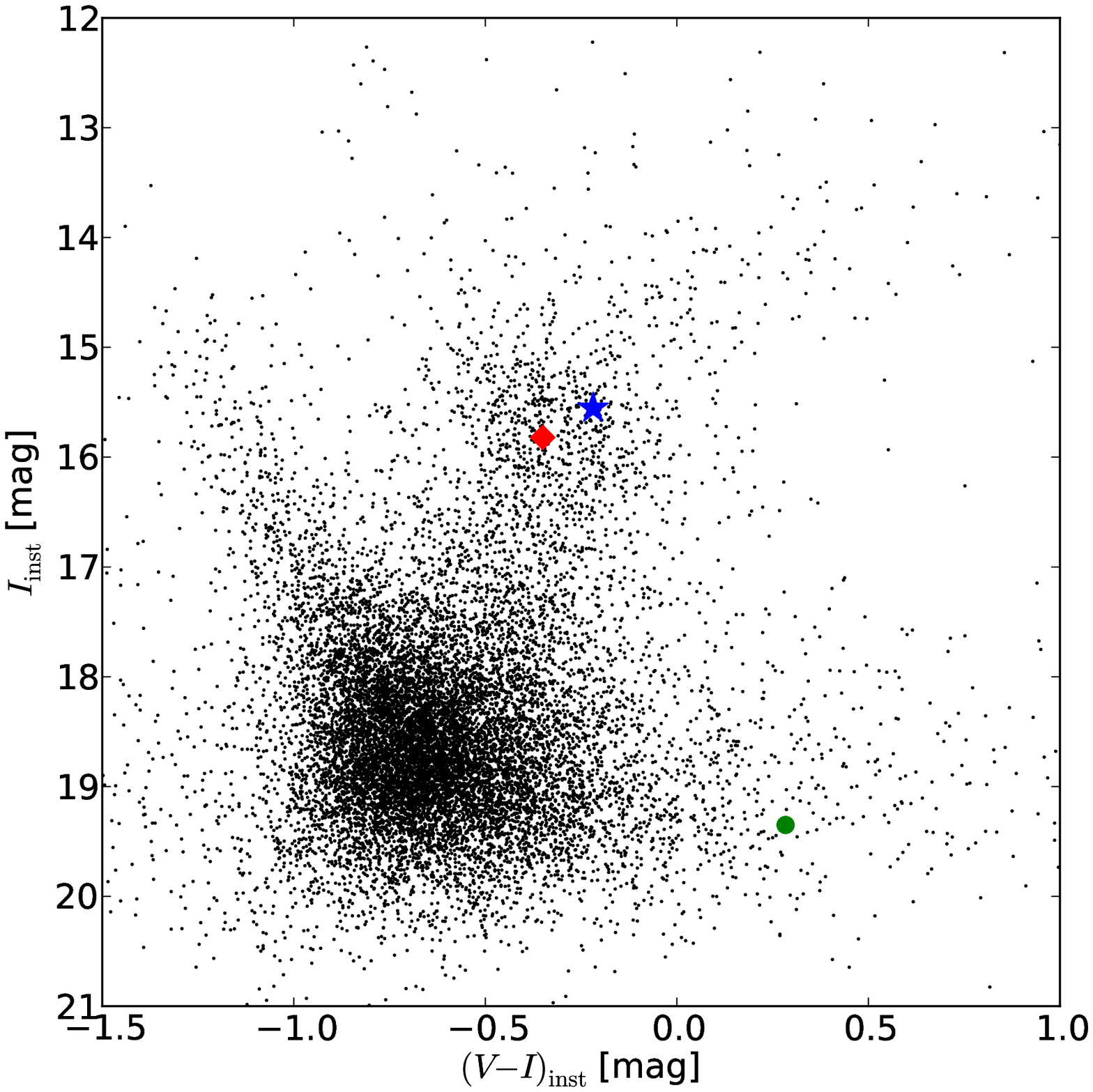}
\caption{The instrumental color-magnitude diagram for the field of view including the
lensing source star.  The position of the source is marked 
with a blue star, relative to the center of the Red Giant Clump highlighted with a red diamond.  The
green circle indicates the color and $V$ magnitude of light blended in the CTIO
photometry. \protect\label{fig:cmd}  }
\end{figure}

Long-exposure $V$, $I$ images were acquired by the CTIO 1.3\,m at several
epochs which enabled us to plot the color-magnitude diagram for the field
including the source star (Fig.~\ref{fig:cmd}).  By observing the event at
different levels of lensing magnification, these data can be incorporated
into the model which yields the source and blended light fluxes, \fluxs,
\fluxb\ for those data, and hence the instrumental magnitudes and colors of
the source and blend.  But we note that these uncalibrated fluxes also
suffer from the high degree of extinction along the line of sight to the
Galactic Bulge.  To calculate the dereddened color, \VISo, and magnitude,
\ISo, of the source, we needed to calibrate the instrumental fluxes \fluxs\
and \fluxb\ relative to a standard candle. 

Fig.~\ref{fig:cmd} clearly shows a locus of stars centered at
\IRCinst=\IRCinstvale\,mag, \VIRCinst=\VIRCinstvale\,mag.  This consists of a
clump of red giant stars, for which stellar theory predicts a stable absolute
luminosity, varying only slightly with age and chemical composition.  Their
frequent occurrence makes these objects useful as standard candles. 
\citet{Stanek1998} established photometric calibrations for red clump magnitudes
which were later refined by \citet{Alves2002} using {\it Hipparcos} data.  Most
recently, \citet{Nataf2012} were able to measure the dereddened apparent
magnitude of the red clump stars at the Galactocentric distance, \IRCo =
\IRCoval.  By mapping the distances, $D_{RC}$, to red clump stars in the
Galactic Bar as a function of Galactic longitude, $l$ they found an apparent
viewing angle on the Bar of $\phi_{\rm Bar}$=40$^{\circ}$, 

\begin{equation}
\frac{R_{0}}{D_{RC}} = \frac{\sin{\phi + l}}{\phi} = \cos{l} + \sin{l}\cot{\phi},
\end{equation}

\noindent where \citet{Nataf2012} measured $R_{0}$ to be \Ro\,kpc. For the
field of MOA-2010-BLG-073 (Galactic coordinates: $l$=4.81030$^{\circ}$,
$b$=-3.50131$^{\circ}$), we derive $D_{RC}$=\DRC\,kpc, and we assumed that the
source star lies behind the same amount of dust as the Red Clump stars and at
the same distance. Scaling the dereddened apparent magnitude of
the red clump stars, \IRCo, appropriately for the slightly closer distance of
the stars in this field, \IRCapp = \IRCo + $\Delta I$, where:

\begin{equation}
\Delta I = 5\log_{10}{R_{0}/D_{RC}}.
\end{equation}

We found $\Delta I$ = \DeltaIval\,mag, and so the distance modulus to the red
clump and the source in this field is \IRCapp = \IRCappval\,mag. 
\citet{Bensby2011} determined the intrinsic \VIRCo=\VIRCoval\,mag for red clump
stars, so we were able to derive their absolute $V$ magnitude of
$M_{V,RC,O}$=\MVRCoval\,mag.  Combining these results with the measured
$\Delta(V-I)_{\rm inst}$ and $\Delta V_{\rm inst}$ between the source and red
clump in the CTIO data, we then derived the dereddened color, \VISo\ and
magnitude \ISo\ of the source, summarized in Table~\ref{tab:physicalpars}. 

\citet{Bessell1988} provided a relationship between $(V-I)$ and $(V-K)$ color
indices, and \citet{Kervella2004} related $(V-K)$ to angular radius for giant and
dwarf stars.  Having thus determined \DS\ and \thetaS, we derived the physical
parameters of the lensing system, which are also summarized in
Table~\ref{tab:physicalpars}.  The color and large source radius of
\RSvale\,\Rsun\ implies this star is a K-type giant, which is consistent with the
observed photometric variability.  \citet{Jorissen1997} found that red giant stars
with spectral types later than early-K are all variable, with amplitudes
increasing from microvariability to several magnitudes towards cooler 
temperatures and timescales from days to years.  \citet{Kiss2006} notes that
irregular photometric variability may be caused by large convection cells, or may
actually be the result of a number of  simultaneous periodic pulsation modes, and
many examples have been identified from time-domain surveys
\citep[e.g.]{Wray2004,Wozniak2004,Eyer2005,Ciechanoska2006}.   We note that the
star was detected by {\sc 2mass} \citep{Skrutskie2006} as source
\objectname[2MASS~J18101138-2631226]{2MASS~J18101138-2631226} with colors
$(J-H)$=0.76$\pm$0.078\,mag and $(H-K_{S})$=0.284$\pm$0.079\,mag.  Although the
{\sc 2mass} field is crowded the star's PSF is distinct and their photometry for
it has the best-quality AAA flag.  These colors are consistent with a giant star
and with our derived value for \KSo=\KSoval\,mag, when we take into account
$AK_{S}$=0.24\,mag from the {\sc vvv} survey \citep{Gonzalez2012}.  

\begin{deluxetable}{lcc}
\tabletypesize{\small}
\tablewidth{7.5cm}
\tablenum{4}
\tablecolumns{3}
\tablecaption{The physical parameters of the lensing system and source star, derived from the model including slope,
parallax and orbital motion,  plus color information.\label{tab:physicalpars}}
\tablehead{\colhead{Parameter} & \colhead{Units} & \colhead{Value}}
\startdata
\thetaS       	      	& $\mu$as     	& \thetaSvale\\
\thetaE       	      	& mas   	& \thetaEvale\\
\RS       	      	& \Rsun       	& \RSvale\\
\MLp     	      	& \Msun       	& \MLpvale\\
\MLs     	      	& \MJ 	      	& \MLsvale\\
\MLtot        	      	& \Msun       	& \MLtotvale\\
\DL      	      	& kpc 	      	& \DLvale\\
$a_{\perp}$   	      	& AU  	      	& \aperpvale\\
KE/PE 	      	      	&     	      	& \Eratio\\
Proper motion 	      	& mas~yr$^{-1}$ & \PM\\[0.2cm]
\ISinst     	      	& mag 	      	& \ISinstvale \\
\VSinst     	      	& mag 	      	& \VSinstvale\\
\VISinst 	      	& mag         	& \VISinstvale\\
\IRCinst    	      	& mag 	      	& \IRCinstvale\\
\VIRCinst     	      	& mag         	& \VIRCinstvale\\
\IRCo    	      	& mag 	      	& \IRCoval \\
\VIRCo        	      	& mag 	      	& \VIRCoval\\
\ISo     	      	& mag 	      	& \ISoval \\
\VSo     	      	& mag 	      	& \VSoval \\
\VISo 	      	        & mag		& \VISovale\\
\VKSo 	      	        & mag		& \VKSoval\\
\KSo 	      	        & mag		& \KSoval\\
$J$ ({\sc 2mass})      	& mag 	      	& \twoMJvale\\
$H$ ({\sc 2mass})      	& mag 	      	& \twoMHvale\\
$K_{S}$ ({\sc 2mass})   & mag 	      	& \twoMKvale\\
\enddata
\end{deluxetable}

\citet{Nataf2012} explain that their value of our viewing angle of the Galactic
Bar is a ``soft upper bound'' because the distance along the plane to the greatest
density of stars along the line-of-sight to a triaxial Bar structure is less than
the distance to the structure's major axis on the far side and greater on the near
side.  As the physical parameters derived for MOA-2012-BLG-073 are somewhat
dependent on the value of our viewing angle of the Galactic Bar and
\citet{Nataf2012} quote consistent results with values as low as 25$^{\circ}$, we
explored the potential impact of this on our results.  A reduced viewing angle
would produce a smaller distance to the source, changing it's dereddened magnitude
and color.  The resulting increase in source radius produces a corresponding
reduction in the value of \DL and increases in the lens masses.  However we found
that the physical parameter values do not change by more than the errors
quoted in Table~\ref{tab:physicalpars}, implying that this is not the dominant
source of uncertainty.  Finally, we computed the physical parameters derived from the best-fitting
\uoneg model for comparison and found that the masses derived changed by $<1\sigma$.  

\section{Discussion}
\protect\label{sec:discussion}

The Earth's movement during this relatively long time scale (\TE=\TEval\,d)
microlensing event resulted in a gradual shift in our perspective on the
lensing system, breaking the symmetry of the caustic.   Meanwhile, the
change in projected separation of the lensing objects modified the shape of
the caustic just as the source's trajectory happened to pass close by.  If
not for these subtle variations, it can be seen from
Figure~\ref{fig:caustic} that a source trajectory \uopos\ would produce
exactly the same light curve as a \uoneg\ trajectory.  As it is, for this
event, the \uopos\  solution best explains our observations, though the
difference in \chisq\ relative to the corresponding  \uoneg\ solution is
very small ($\Delta\chi^{2}$=20.2) compared with the \chisq\ of both fits.  

The measurable parallax signature enables us to determine the masses of the
lensing bodies.  The primary lensing object has \MLp=\MLpvale\,\Msun, making
it an M-dwarf star.  The companion's mass is \MLs=\MLsvale\,\MJ.  This
places it the brown dwarf desert, though we note that this traditionally
refers to close-in companions, and since microlensing and direct imaging
measure only the projected separation, we know only their {\it minimum}
orbital semi-major radii.   Regardless, it is clear that
MOA-2010-BLG-073L\,b is close to the mass threshold for deuterium burning
($\sim$0.012\,\Msun=12.6\,\MJ) quoted as the nominal boundary between
planets and brown dwarfs established by the IAU \citep{Chabrier2005}.  So
what kind of object is it?  \\


\begin{figure}
\figurenum{8}
\epsscale{1.0}
\plotone{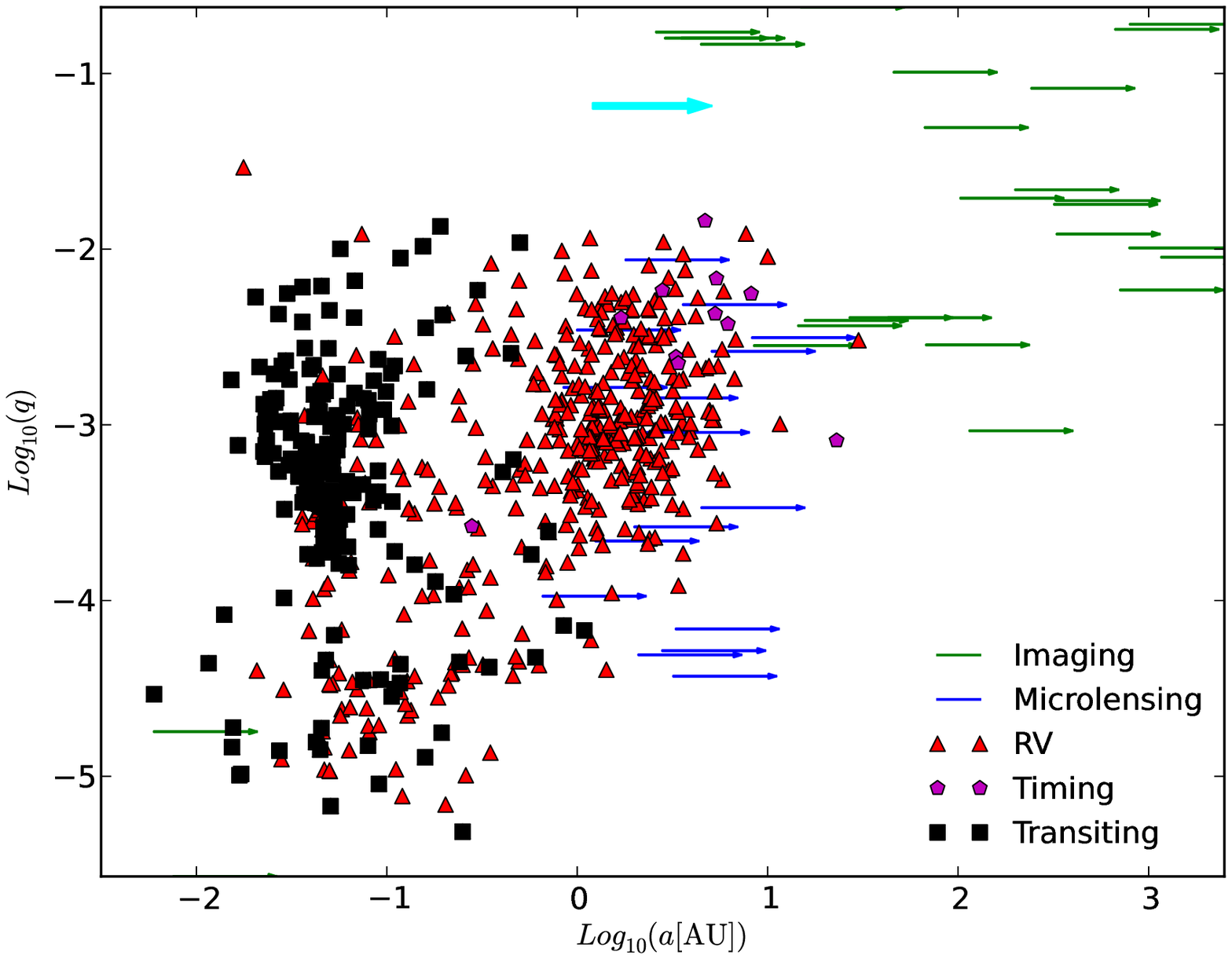}
\caption{The ratio of planet mass to host star mass plotted against semi-major
axis for all exoplanets for which these parameters are available (source:
exoplanet.eu \cite{Schneider2011}).  The true orbital semi-major axis is
plotted where it has been measured, otherwise the projected  (minimum)
separation has been indicated with an arrow, where the base of the arrow marks
the measured value.  Very few objects have been found between $\log_{10}(q)$=-1 --
-2.  This corresponds to the brown dwarf desert.  The location of
MOA-2010-BLG-073L, highlighted in cyan and bold, lies between the brown dwarf
and planetary regimes.  \protect\label{fig:massratio}  }
\end{figure}


No further orbital or metallicity information is available for this event,
which might have shed light on its evolutionary history.   Theoretical
isochrones predict that a star of this low mass will not have lost a
significant amount of material over its lifetime, so we can say that
MOA-2010-BLG-073L\,b formed as a high mass-ratio binary.  Models of
protoplanetary disks have the expectation that disk mass, $M_{D}$, will
scale linearly with star mass $M_{*}$ \citep{Williams2011},
$M_{D}/M_{*}\sim1$\% at young ages, but this would limit $M_{D}$ -- and
hence $M_{P,\rm{max}}$ -- to $\sim$1--10\,\MJ\ in the case of
MOA-2010-BLG-073L.  So it seems questionable whether such a massive companion
could have formed in a protostellar disk, via either core accretion or gravitational 
instability.  

\citet{Bonnell2008,Kroupa2003} discuss a number of mechanisms which can
produce an M-dwarf/Brown Dwarf binary following gravitational fragmentation
in a molecular cloud:

\begin{enumerate}
\item {\em Embryo rejection model} the nascent binary was ejected from a
dynamically unstable multiple protostellar system, leading to the loss of its
accretion envelope before the secondary component could acquire enough mass to
become a star.  
\item {\em Collision model} the binary was prematurely ejected from a larger
protostellar system by the close passage of another star.
\item {\em Photo-evaporation model} the accretion envelope around the binary was
photo-evaporated by the nearby presence of a massive star in the birth cluster
before the secondary could accrete enough mass to become a star. 
\item {\em ``Star-like'' model} the object formed as a normal stellar binary
with low-mass components.  
\end{enumerate}

The embryo rejection model predicts \citep{Bate2002} that the maximum separation
of binaries surviving this process is $a_{\rm{max}}\sim10-20$\,AU.  We cannot rule
out this scenario as we only measure the projected separation of the lens, which
is nevertheless below $a_{\rm{max}}$.  Since MOA-2010-BLG-073L is a field object
we have no information regarding the proximity of other stars during its birth, so
the collisional and photo-evapouration models are equally plausible.  However, we
note that \citet{Whitworth2006} derived a minimum mass for primary fragmentation
components of 0.004\,\Msun$\sim$4.2\,\MJ.  As this threshold is below the mass of
MOA-2010-BLG-073Lb the simplest explanation is that the companion is an extremely
low-mass product of star formation.  However, we note that the mass ratio of this
system would match that of an object in the brown dwarf desert if the host were a
more massive star.  Recent surveys (\citet[e.g.]{Dieterich2012, Evans2012},
though restricted to massive companions with $a>10$\,AU) hint
that the brown dwarf desert may extend to M-dwarfs and beyond 3\,AU, which would
make MOA-2010-BLG-073L a rare example of its type. 

\acknowledgments

RAS is grateful to the Chungbuk University group and especially to C.~Han and
J.-Y.~Choi for their advice and hospitality in Korea where much of this work was
completed.  RAS would also like to express appreciation for Y.~Tsapras,
K.~Horne, M. Hundertmark, S.~Dong and P.~Fouqu{\'e} for many useful
discussions.   KA, DMB, MD, KH, MH, CL, CS, RAS and YT would like to thank the
Qatar Foundation for support from QNRF grant NPRP-09-476-1-078.  Work by C.~Han
was supported by the Creative Research Initiative Program (2009-0081561) of the
National Research Foundation of Korea.  CS received funding from the European
Union Seventh Framework Programme (FP7/2007-2013) under grant agreement
no.~268421. The OGLE project has received funding from the European Research
Council under the European Community's Seventh Framework Programme
(FP7/2007-2013) / ERC grant agreement no. 246678 to AU.  This publication makes
use of data products from the Exoplanet Encyclopeida and the Two Micron All Sky
Survey, which is a joint project of the University of Massachusetts and the
Infrared Processing and Analysis Center/California Institute of Technology,
funded by the National Aeronautics and Space Administration and the National
Science Foundation. MOA acknowledge funding JSPS20340052, JSPS22403003 and
JSPS19340058.  TCH
gratefully acknowledges financial support from the Korea Research Council for
Fundamental Science and Technology (KRCF) through the Young Research Scientist
Fellowship Program. CUL and TCH acknowledge financial support from KASI
(Korea Astronomy and Space Science Institute) grant number 2012-1-410-02.
DR (boursier FRIA), FF (boursier ARC) and J.~Surdej acknowledge support from the
Communaut\'e francaise de Belgique - Actions de recherche concert\'ees -
Acad\'emie universitaire Wallonie-Europe.  A.~Gould acknowledges support from
NSF AST-1103471.  B.S.~Gaudi, A.~Gould and R.W.~Pogge acknowledge support from
NASA grant NNX12AB99G.  Work by J.C.~Yee is supported by a National Science
Foundation Graduate Research Fellowship under Grant No.~2009068160. 
S.D. is supported through a Ralph E. and Doris M. Hansmann Membership
at the IAS and NSF grant AST-0807444.

{\it Facilities:} \facility{FTN}, \facility{FTN}, \facility{Liverpool: 2m},
\facility{CTIO: 1.3m}, \facility{MtCanopus: 1m}, \facility{SAAO: 1m},
\facility{Danish 1.54m Telescope}

\clearpage

\end{document}